\documentclass[conference]{IEEEtran}
\IEEEoverridecommandlockouts

\usepackage{cite}
\usepackage{amsmath,amssymb,amsfonts}
\usepackage{algorithmic}
\usepackage{graphicx}
\usepackage{textcomp}
\usepackage{xcolor}
\usepackage{url}
\usepackage{subcaption}
\usepackage{booktabs}
\usepackage[linesnumbered,ruled,vlined]{algorithm2e}
\def\BibTeX{{\rm B\kern-.05em{\sc i\kern-.025em b}\kern-.08em
    T\kern-.1667em\lower.7ex\hbox{E}\kern-.125emX}}

\begin{document}

\title{Efficient Cloud-Edge-Device Query Execution Based on Collaborative Scan Operator}

\author{

    \IEEEauthorblockN{1\textsuperscript{st} Chunyu Zhao, 2\textsuperscript{nd} Hongzhi Wang\IEEEauthorrefmark{1}, 3\textsuperscript{rd} Kaixin Zhang, 4\textsuperscript{th} Hongliang Li, 5\textsuperscript{th} Yihan Zhang,\\ 6\textsuperscript{th} Jiawei Zhang, 7\textsuperscript{th} Kunkai Gu, 8\textsuperscript{th} Yuan Tian, 9\textsuperscript{th} Xiangdong Huang, 10\textsuperscript{th} Jingyi Xu}
    \IEEEcompsocitemizethanks{\IEEEcompsocthanksitem\IEEEauthorrefmark{1}Corresponding author.}
    \IEEEauthorblockA{
    \textit{Massive Data Computing Lab, Harbin Institute of Technology}
    \\24B903032@stu.hit.edu.cn, wangzh@hit.edu.cn, {21B903037, 2022111894, 2021110515, 2021113150, 24S003033}@stu.hit.edu.cn }
    \IEEEauthorblockA{
    \textit{Timecho Ltd, Tsinghua University, Advanced Institute of Big Data}
    \\yuan.tian@timecho.com, huangxdong@tsinghua.edu.cn, xujy02@126.com
    }
}

\maketitle

\begin{abstract}
In cloud-edge-device (CED) collaborative query (CQ) processing, by leveraging CED collaboration, the advantages of both cloud computing and edge resources can be fully integrated. However, it is difficult to implement collaborative operators that can flexibly switch between the cloud and the edge during query execution. Thus, in this paper, we aim to improve the query performance when the edge resources reach a bottleneck. To achieve seamless switching of query execution between the cloud and edge, we propose a CQ processing method by establishing a CED collaborative framework based on the collaborative scan operator, so that query execution can be transferred to the cloud at any time when the edge resources are saturated. Extensive experiments show that, under sufficient network download bandwidth, the CED collaborative scan operator can effectively alleviate the performance degradation of scan operators caused by high I/O load and CPU wait time at the edge. It also achieves balanced resource scheduling between the cloud and edge.
\end{abstract}


\section{Introduction}
\begin{figure}[h!]
    \setlength{\belowcaptionskip}{-20pt}
    \centering
    \includegraphics[scale=0.66]{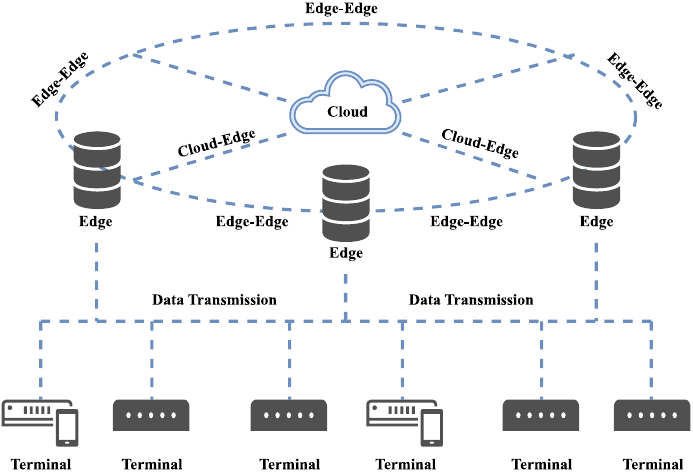}
    \caption{Architecture of CED collaboration}
    \label{fig:1-1}
\end{figure}

CED-CQ distributes query tasks originally handled by a single cloud server with the edge and device sides in a coordinated manner, enabling cloud servers and edge computers to work together to process query tasks efficiently, as shown in Figure~\ref{fig:1-1}. The design of basic CQ operators for CED databases not only facilitates CQs between cloud servers and edge computers but also alleviates the disadvantages of transmission pressure and insufficient computing power in traditional cloud-edge cooperation~\cite{b2}. This complementary approach leverages the advantages of both, offering the convenience of data access with minimal network load while providing ample computing resources and effective coordination~\cite{b3,b4}. In the field of Internet of Things (IoT)~\cite{b5}, due to the large number of devices, high frequency of data generation, and the predominance of time as the primary dimension, time-series databases~\cite{b6} have gradually become the mainstream choice for CED Databases~\cite{b7}.

In the CED database, data collected by terminal devices is initially stored at the edge computer. Meanwhile, a cache is configured on the cloud server to store a portion of key or frequently accessed data. When executing queries, the edge side typically retrieve data directly from local storage. To enhance data availability, the edge side periodically synchronize the most recently queried relevant data to the cloud cache during network idle periods. However, when edge sides experience degraded performance due to high volumes of concurrent queries, they offload part of the query workload to the cloud server. This mechanism not only effectively reduces the load on edge sides, but also leverages the abundant computational resources available in the cloud.

However, enabling efficient CED-CQ processing presents several key challenges:

\begin{itemize}
\item \textbf{Coordinated Data Management:} The cloud and edge must efficiently manage distributed datasets in a coordinated manner. This requires maintaining data consistency while supporting flexible query execution across both cloud and edge.

\item \textbf{Cross-Tier Operator Collaboration and Switching:} Query operators need to be dynamically deployable and executable across cloud and edge tiers. They must also support seamless transitions between execution tiers based on runtime conditions.

\item \textbf{Execution State Synchronization and Migration:} To support seamless operator transitions, the execution state and progress information must be synchronized or rapidly migrated between the cloud and edge.

\item \textbf{Cost-Efficient Resource and Data Transfer Optimization:} It is essential to minimize cross-tier data transfer costs while jointly optimizing compute and storage resource scheduling across the cloud and edge to balance overall performance and cost-efficiency.
\end{itemize}

Even though some techniques have been proposed for CED-CQ processing, they fail to address these issues. 

For instance, the time-series database InfluxDB~\cite{b8,b9} has been applied to assist in data query needs between the cloud and edge. Similarly, one of the key factors contributing to the widespread adoption of Apache IoTDB ~\cite{b10}\cite{b11}\cite{b12} in domains such as connected vehicles, intelligent operations and maintenance, and smart factories is its efficient collaborative data management mechanism~\cite{b13}, which enables seamless synchronization of data across cloud, edge, and devices through the pipe~\cite{b14}.

However, in existing systems, query execution across the cloud and edge remains largely disjoint. The core limitation lies in the absence of a unified query optimization framework capable of integrating and managing data across cloud–edge environments. Representative systems such as InfluxDB and Apache IoTDB treat the cloud and edge as isolated entities, executing queries independently within their respective domains. This fragmented architecture leads to significant inefficiencies, including redundant data transfers and excessive load on edge nodes. More critically, the lack of synchronization in query execution progress between cloud and edge becomes a major bottleneck, severely hindering coordination. As a result, it becomes extremely difficult to build a truly seamless and efficient collaborative query processing mechanism. These limitations significantly restrict the potential of existing systems to support practical and performant CED-CQ processing.

The development of an efficient CED-CQ operators is pivotal for enabling true coordination between cloud and edge in query execution, overcoming the independent processing paradigm that currently limits systems. Unlike traditional approaches where the cloud and edge operate in silos, a well-designed CED-CQ operator can act as a unifying layer that integrates their capabilities into a cohesive query execution framework.

Among all data operators in time series databases, we observe that the scan operator~\cite{b15} plays the crucial role. 

First, a query plan consists of multiple operators, and the scan operator is typically the starting point of these operators, which determines the initial state of the entire query execution process and is often the most time-consuming part of the query.

Second, the scan operator serves as the foundation for many other database operations by providing the initial data required for processing. For instance, during join operations, the efficiency of the join largely depends on the scan operator's ability to retrieve relevant data from two tables effectively. Similarly, in aggregation operations, such as calculating averages or sums over time series, the scan operator filters and reads the relevant data subsets, reducing computational overhead. Additionally, in sorting and group-by operations, the scan operator ensures that only the necessary data is accessed and passed on, which helps streamline the organization or categorization of results. In summary, the scan operator forms the foundation of query processing, which enables the implementation of core operations such as aggregation, selection, projection, and join.

\begin{figure}[h!]
    \setlength{\belowcaptionskip}{-5pt}
    \centering
    \includegraphics[scale=0.35]{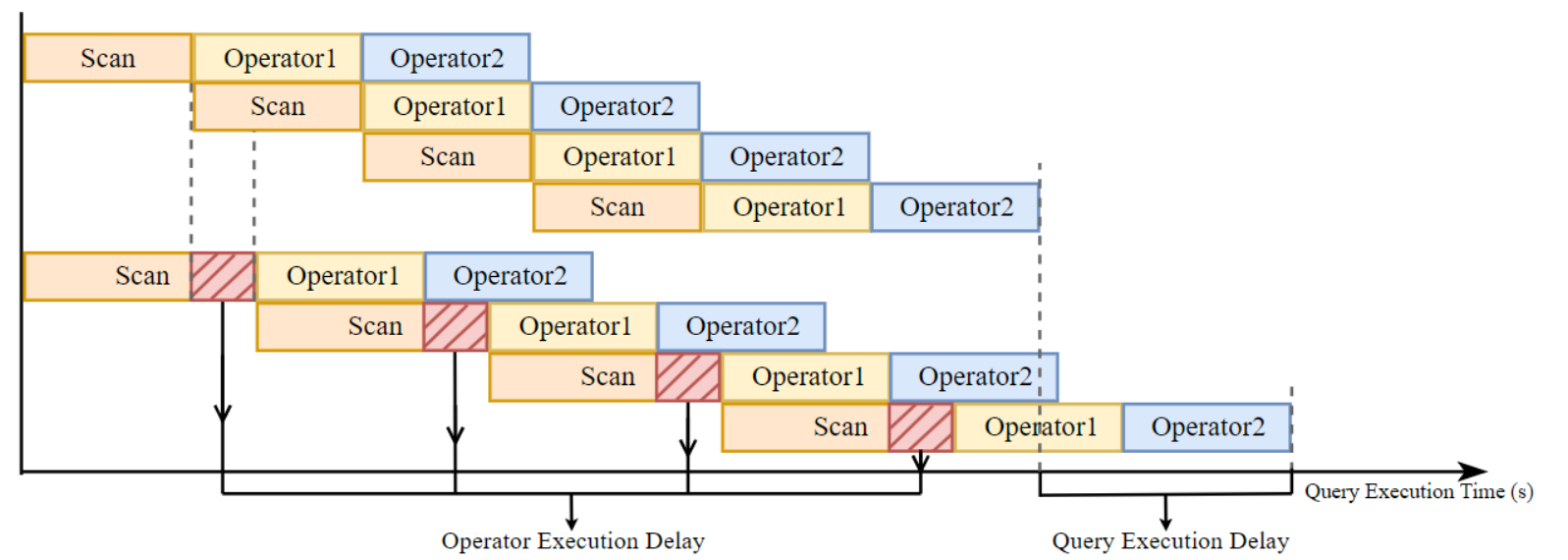}
    \caption{Impact of resource overload on operator execution}
    \label{fig:1-2}
\end{figure}

The last but not least, when I/O overload occurs, the scan operator can impact the efficiency of all other operators. Due to time-series database's use of the Volcano model~\cite{b16}, each operator returns query results in the \textit{next()} method, forming a pipeline. When the read rate of the scan operator, which acts as a leaf node, decreases, its parent node will continuously wait for the scan operator's return results. Consequently, this bottom-up delay affects that of all operators. Therefore, if the execution time of the parent node's operator is less than that of the scan operator, it will continuously accumulate the additional time required for the scan operator to read, leading to a decrease in query efficiency. The specific process is illustrated in Figure \ref{fig:1-2}.
\vspace{-10pt}

Thus, to address the challenges of CED-CQ processing, we focus on the CED scan operator capable of flexible switching between the cloud and the edge. 

Compared to the cloud server, the edge in a CED database typically has more limited computation and storage resources, making it more susceptible to overload. Since the edge sides often rely on mechanical hard drives, I/O performance becomes one of the dominant factors affecting query execution. Under high I/O pressure, the CED database currently lacks the capability to dynamically offload ongoing query tasks to the cloud. As a result, the edge may experience a series of issues, including excessive I/O contention, slow data scans, and increased query latency due to bottlenecks in the scan operator.

In addition to edge I/O overload, CPU resource contention caused by a surge of concurrent queries is another key factor contributing to edge congestion. When the volume of concurrent queries exceeds the computational capacity of the edge, CPU scheduling latency increases exponentially, leading to query queue buildup and degraded response times.

Therefore, exploring and developing an efficient collaborative scan operator that seamlessly integrates cloud, edge and device resources to achieve efficient data collaboration processing and cloud-edge collaborative queries has become a critical issue.

To enable efficient CED Scan, the core challenge lies in establishing a real-time cross-tier coordination mechanism that allows the cloud and edge to dynamically cooperate during query execution. This requires precise coordination of data access paths between tiers, ensuring that distributed scan operators can support collaborative execution with low-overhead state interaction and millisecond-level progress synchronization. Specifically, this challenge can be decomposed into the following key issues:

\begin{enumerate}
\item \textbf{Cross-Tier Data State Consistency:} Building unified data management across cloud and edge tiers requires addressing challenges such as dynamic index synchronization, cache coherence maintenance, and distributed data positioning. The key objective is to ensure that all compute nodes access logically consistent datasets, eliminating redundant transfers or execution conflicts caused by inconsistent data placement awareness.

\item \textbf{Collaborative Scan Operator Architecture:} A distributed scan operator must be designed to support dynamic task migration. Key challenges include: (i) adaptive data transmission pipe mechanisms that select appropriate transmission modes based on query characteristics; (ii) pipe creation and isolation in multi-scan-operators scenarios;  (iii) ensuring consistent query parsing and execution semantics across tiers.

\item \textbf{Delta State Synchronization Protocol:} To enable seamless operator relocation between cloud and edge, a lightweight delta state synchronization mechanism is required. This includes: (i) real-time exchange of lightweight execution progress; (ii) asynchronous waiting compensation strategy to tolerate network variability; (iii) timely consistency guarantees for delta state updates.
\end{enumerate}

To tackle these challenges, we propose a collaborative execution framework based on dynamic load migration. The core idea is to mitigate performance bottlenecks in CED-CQ processing, especially under edge resource saturation, through runtime coordination and elastic task redirection.

Specifically, we design a family of adaptive collaborative scan operators, including the series scan operator and aggregation scan operator, both of which support runtime data path redirection. When I/O or CPU congestion is detected at the edge, the operator seamlessly switches its data source from local storage to a cloud-based streaming interface, enabling on-the-fly migration of query workloads from edge to cloud.

To enable this dynamic behavior, we develop an elastic load scheduling engine. This engine monitors real-time edge resource utilization and triggers fine-grained operator migration to the cloud when predefined thresholds are exceeded. Once resource pressure subsides, the engine supports operator fallback to the edge, forming a closed-loop, bidirectional control mechanism that balances responsiveness and resource efficiency.

Furthermore, we implement a distributed CQ transmission pipe that supports incremental state synchronization and batch-optimized data transfer across cloud and edge. To ensure throughput stability under high load, we introduce techniques such as prefetching and pipe reuse.

Through this architecture, the three key challenges are addressed in a unified manner, laying the foundation for scalable, real-time CED-CQ execution.

\vspace{-12pt}
\begin{figure}[h!]
    \setlength{\belowcaptionskip}{-10pt}
    \centering
    \includegraphics[scale=0.53]{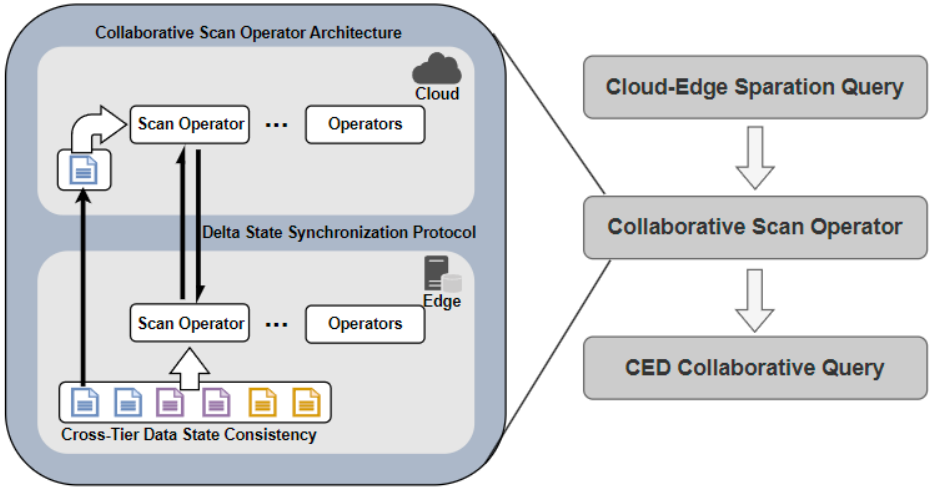}
    \caption{Main research content framework}
    \label{fig:1-3}
\end{figure}

The main contributions of this paper are as follows:
\begin{itemize}
    \item We construct a cross-tier CED-CQ architecture to enable dynamic data transfer and operator migration between the cloud and edge. This design effectively addresses the fragmented execution problem in traditional architectures.
    
    \item We design collaborative scan operators including series and aggregation scan operators, along with their adaptive coordination algorithm, which achieves dynamic balance between the edge resource pressure and the cloud computation through runtime resource monitoring and closed-loop cloud resource scheduling.
    
    \item We develop a fine-grained resource scheduler tailored for collaborative scanning, supporting seamless operator state migration and elastic resource allocation, ensuring cloud-to-edge execution continuity under high-load conditions.

    \item Extensive experiments validate the effectiveness of the CED-CQ under typical bandwidth conditions, which is $>$ 50 Mbps. The results show that it reduces the scan latency by 62\%–79\% under high I/O pressure and 62\%–79\% under high CPU pressure compared to the baseline systems, significantly alleviating the read bottleneck of the edge.
\end{itemize}

In Section~\ref{sec2}, we review the current research landscape of CED queries and discuss their advantages and limitations. In Section~\ref{sec3}, we overview the CED databases and introduce the dynamic execution framework based on collaborative scan operators. In Section~\ref{sec4}, we describe the load-aware operator coordination algorithm and the mechanism for maintaining cross-tier data state consistency. In Section~\ref{sec5}, we provide a detailed design of the collaborative scan operator architecture. In Section~\ref{sec6}, we introduce the incremental state synchronization protocol for seamless operator migration. In Section~\ref{sec7}, we conduct comparative experiments to demonstrate how the CED-CQ effectively mitigates scan operator slowdown caused by resource saturation. In Section~\ref{sec8}, we show how the CQ pipeline improves the edge's query performance under high resource saturation, discuss the limitations of our proposed design, and outline potential directions for future work.

\section{Related Work}\label{sec2}

\subsection{CED data storage}\label{subsec2.1}
The current methods for CED data storage mainly follow two strategies: centralized and distributed. Literature~\cite{b17} proposes a method for periodically collecting data from sensor nodes and sending it to a centralized database. However, this can create a bottleneck on the server due to the large volume of data generated~\cite{b18}. Kanzaki et al. discussed a wireless sensor network test platform, X-Sensor, based on a centralized approach~\cite{b19}
. This system stores data collected from all nodes in a centralized database. Elias et al. proposed a centralized storage-based monitoring method for wireless sensor networks~\cite{b20}, which collects sensor data and stores it in a centralized relational database, accessed by mobile clients.

Although distributed databases~\cite{b21} can be used to manage data in sensor networks, traditional distributed databases are not suitable for large-scale sensor networks because they require metadata maintenance~\cite{b22}. While metadata can be maintained in nodes via static links in wireless sensor networks, ensuring metadata consistency is challenging due to link and node failures.

In terms of databases supporting CED environments, OpenTSDB~\cite{b23} and HBase~\cite{b24} only support the cloud data storage. InfluxDB~\cite{b25} supports both the cloud and the edge data storage. TDengine~\cite{b26}, Machbase~\cite{b27}, and Apache IoTDB support data storage across cloud, edge, and terminal, but the first two do not support collaborative processing; only the latter supports collaborative management of cloud-edge data.

Although research on CED data storage has begun, current storage methods mainly focus on single the cloud, edge, or device storage. Applying these directly to CED collaborative scenarios presents two significant drawbacks. Firstly, they do not account for the heterogeneous characteristics of CED devices and data, making it difficult to meet the real-time and availability requirements for collaborative storage of high-dimensional time-series data in CED coordination. Secondly, they struggle to directly support collaborative data operations in terms of storage structure, failing to meet the needs for computational power sinking to the edge in CED collaboration.

\subsection{CED collaborative query}\label{subsec2.2}
With the rise of cloud computing and edge computing, some emerging databases are better suited to meet the needs of CED collaborative queries. For example, the time-series database InfluxDB supports collaborative data management across cloud, edge, and terminal, but its performance in supporting cloud-edge data queries is somewhat lacking. Apache IoTDB, although it supports data management on CED~\cite{b28}, only offers limited rule-based query optimization techniques and separate query execution~\cite{b29} for cloud and edge, falling short in achieving CQ processing and optimization across CED.

Although there are currently systems that support IoT data queries, most of these solutions only cover one or two layers among cloud, edge, and device. Even the IoTDB, which supports CED data management, fails to achieve CQ processing and optimization across these layers. The root cause of these issues lies in the nascent stage of research on CED collaborative data management systems. Most existing systems lack foundational support for collaborative queries across CED in their underlying design, particularly in basic data operations, CQ optimization methods, and processing techniques for CED collaboration.

\section{System Overview}\label{sec3}

In this section, we outline the foundational assumptions underlying our system and provide a comprehensive overview of its system and the dynamic execution framework based on collaborative scan operators.
\subsection{System Assumptions}\label{subsec3.1}
To establish a clear system definition for CED-CQ and to focus on operator design constraints, we define the following core assumptions:

\begin{enumerate}
\item \textbf{Hierarchical Deployment Architecture:}
\begin{itemize}
\item \textit{Cloud Tier:} Includes at least one \textit{ConfigNode}\footnote{Manage cluster's node information, configuration information, etc., responsible for scheduling and balancing distributed operations.} and $k$ \textit{DataNodes}\footnote{Serve client requests, responsible for storing and computing data.}, where $k \geq 1$.
\item \textit{Edge Tier:} Includes at least one cluster. Each cluster is Composed of $m$ edge computers, a cluster contains:
\begin{itemize}
\item A pool of \textit{ConfigNodes}: $n_{\text{cfg}} \geq 1$
\item A pool of \textit{DataNodes}: $n_{\text{data}} \geq 2$ (to meet minimum replication constraints)
\end{itemize}
\end{itemize}
\item \textbf{Cloud Resource Abundance:}
Cloud computing resources are assumed to be elastic and effectively infinite.

\item \textbf{Distributed Data Storage Strategy:}
\begin{itemize}
\item \textit{Primary Storage Layer:} Edge \textit{DataNodes} store raw data using partitioning rules~\cite{b13}, satisfying:

$Data = \prod_{i=1}^{n} d_i \in DataNode_i$

\item \textit{Cloud Cache Layer:} The cloud caches a frequently accessed \textit{DataNode} data from edge data using the least recently used (LRU) policy~\cite{b38}:

$$Cache_{\text{cloud}} = \{ d | freq(d) > \tau_{\text{hot}} \land d \in \bigcup_i DataNode_i \}$$

The $freq()$ refers to frequency of access to data item. The $\tau_{\text{hot}}$ refers to the threshold in the LRU policy.  
\end{itemize}

\item \textbf{Industrial IoT Query Pattern:}
Due to physical deployment constraints in industrial environments, queries are typically initiated from edge clients.
\end{enumerate}

\subsection{Dynamic Execution Framework in CED Database}\label{subsec3.2}

We begin by introducing the hierarchical architecture design of the CED  database. The current CED system adopts a tiered deployment structure composed of a central cloud server and multiple distributed edge clusters. The cloud layer hosts one \textit{ConfigNode} and at least one \textit{DataNode}, while each edge cluster contains at least one \textit{ConfigNode} and several \textit{DataNodes}.

\begin{figure}[h!]\
    \setlength{\belowcaptionskip}{-10pt}
    \centering
    \includegraphics[scale=0.4]{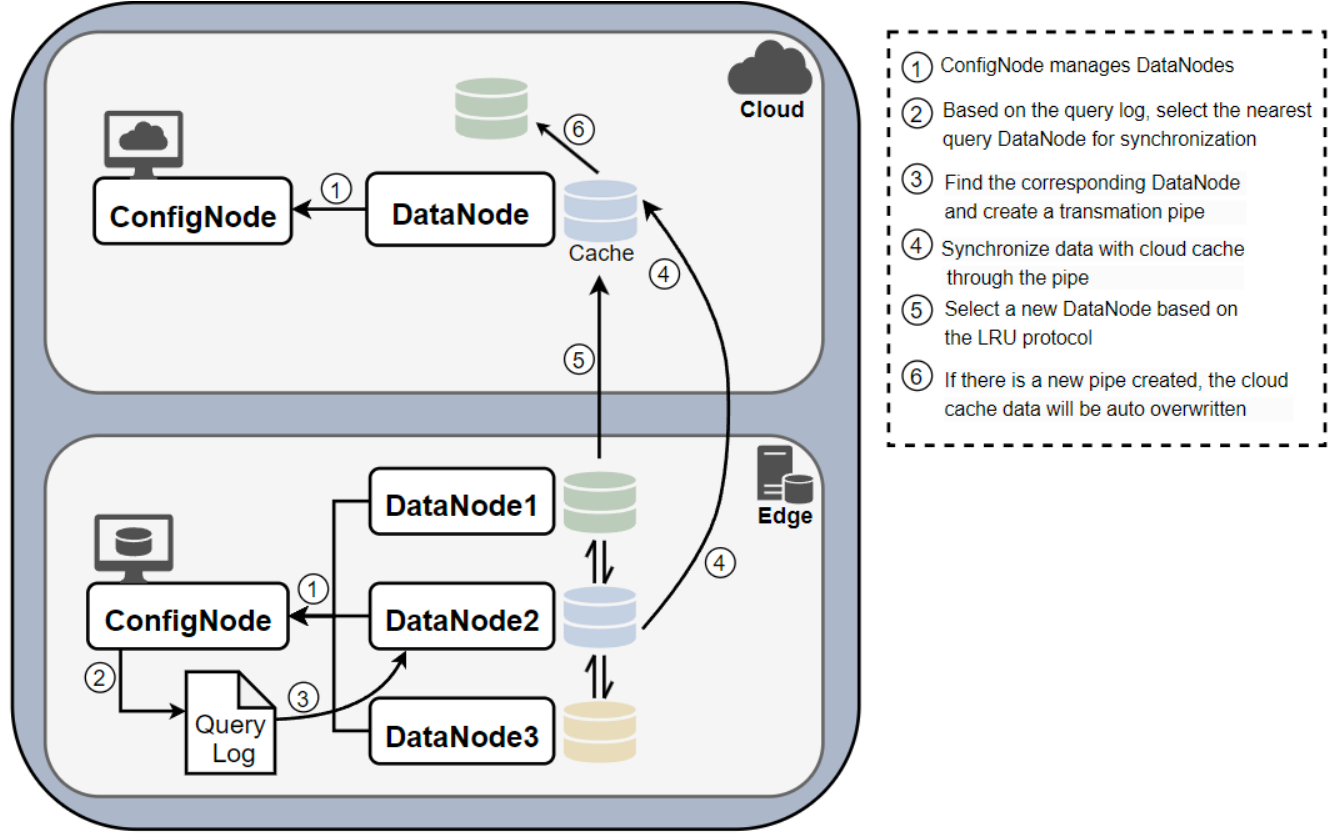}
        \caption{Workflow of the cloud cache}
    \label{fig:3-1}
\end{figure}

Data collected by the devices, such as sensors, are first persistently stored in the edge \textit{DataNode}s. Within each edge cluster, cross-node data replication is achieved through the raft consensus protocol~\cite{b13}. Each \textit{DataNode} primarily stores data from its associated sensors and maintains only limited replicas of data from other \textit{DataNode}s. As shown in Figure~\ref{fig:3-1}, to optimize query performance, when bandwidth utilization remains below a predefined threshold, the CED database leverages a LRU policy to synchronize frequently accessed data from hotspot edge \textit{DataNode}s to the cloud cache layer.

We now describe the query execution workflow, which follows a dynamic load-aware scheduling principle. When a query is issued by an edge client, the system monitors resource utilization in real time. If no resource bottlenecks are detected, the query is executed directly by the local \textit{DataNode} associated with the client. The system adopts a streaming result mechanism that transfers data in batches using fixed-size units called \textit{TsBlocks}~\cite{b32}, each containing 1,000 records. For instance, when the result set exceeds 1,000 records, the first \textit{TsBlock} is returned immediately, and subsequent blocks are transmitted on demand when explicitly requested by the client. If the required data are not locally available on the executing \textit{DataNode}, the query operator will automatically initiate cross-\textit{DataNode} fetching, enabling transparent intra-cluster data access.

\begin{figure}[h!]
    \setlength{\belowcaptionskip}{-20pt}
    \centering
    \includegraphics[scale=0.3]{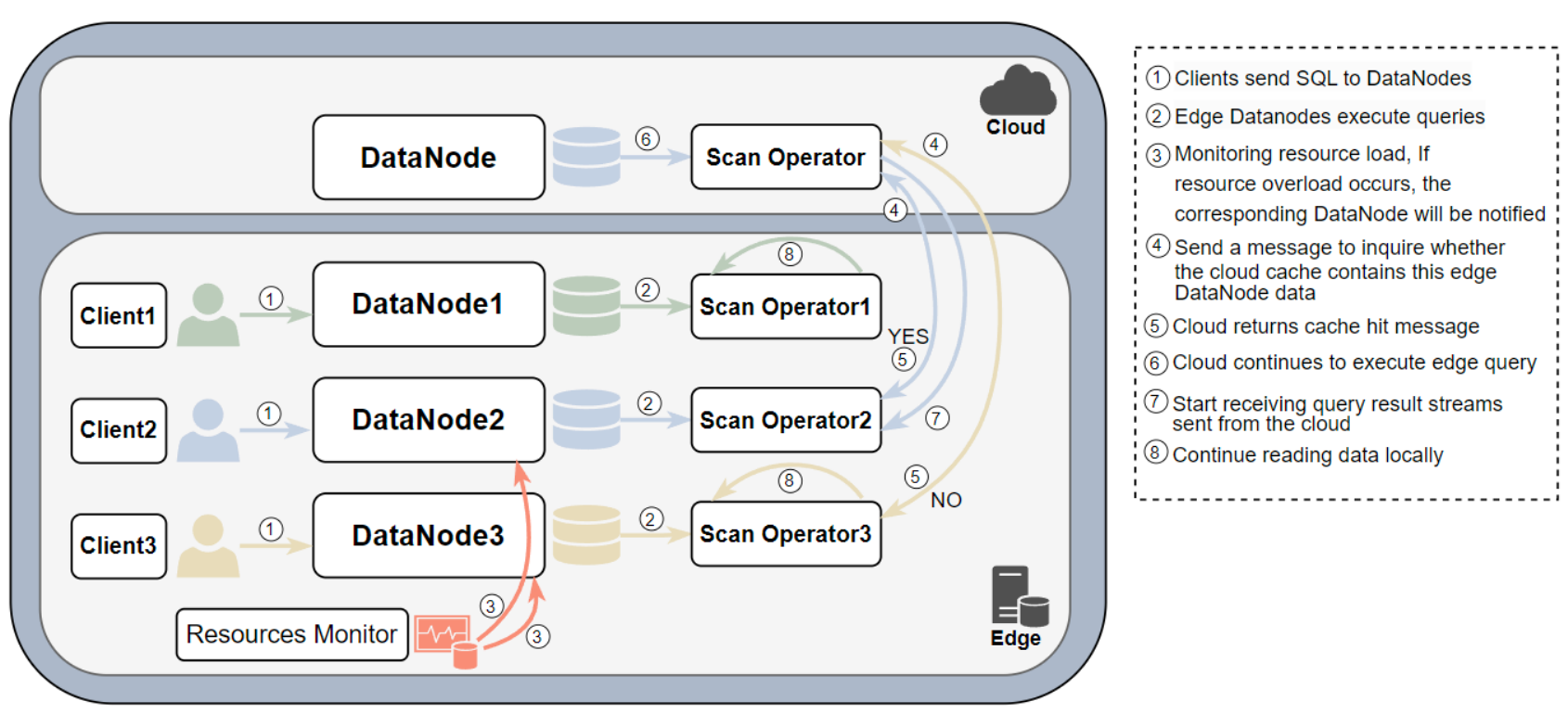}
    \caption{Overview of the CED CQ execution}
    \label{fig:3-2}
\end{figure}

As shown in Figure \ref{fig:3-2}, when the resource monitoring module detects overload, the CED cooperative execution mechanism is triggered. The edge \textit{DataNode} encapsulates the current operator execution state and incremental delta into a migration package and sends it to the cloud. Upon verifying that the required data is available in the cloud cache and the takeover conditions are satisfied, the cloud returns a confirmation signal. The edge \textit{DataNode} then suspends local execution and redirects data access to the cloud stream. The cloud resumes the query execution from the interruption point using the received state and delta, returning results in the form of \textit{TsBlocks} to the edge \textit{DataNode}, thus achieving seamless operator migration across layers.

In addition, the system ensures adaptive resource balancing through continuous monitoring. When the edge resource utilization falls back to a safe threshold and remains stable, the system initiates a remigration process. The cloud terminates the current task and sends a final delta package to the edge. The edge \textit{DataNode} then reconstructs the operator context and resumes query execution locally. The data source is automatically switched back to the local store, ensuring uninterrupted query processing without requiring any client intervention.

\section{Cross-Tier Data State Consistency}\label{sec4}

To enable seamless dynamic task migration, it is essential to establish a cross-tier data state consistency mechanism that ensures both cloud and edge access a logically unified dataset. This eliminates redundant queries or execution conflicts caused by inconsistent state perception. Achieving this goal requires three key capabilities:(i) efficient distributed data localization, (ii) real-time consistency maintenance between cloud cache and edge, (iii) and a collaborative synchronization protocol for dynamic data indexing.

\subsection{efficient distributed data localization}\label{subsec4.1}

\begin{figure}[h!]
    \setlength{\belowcaptionskip}{-20pt}
    \centering
    \includegraphics[scale=0.28]{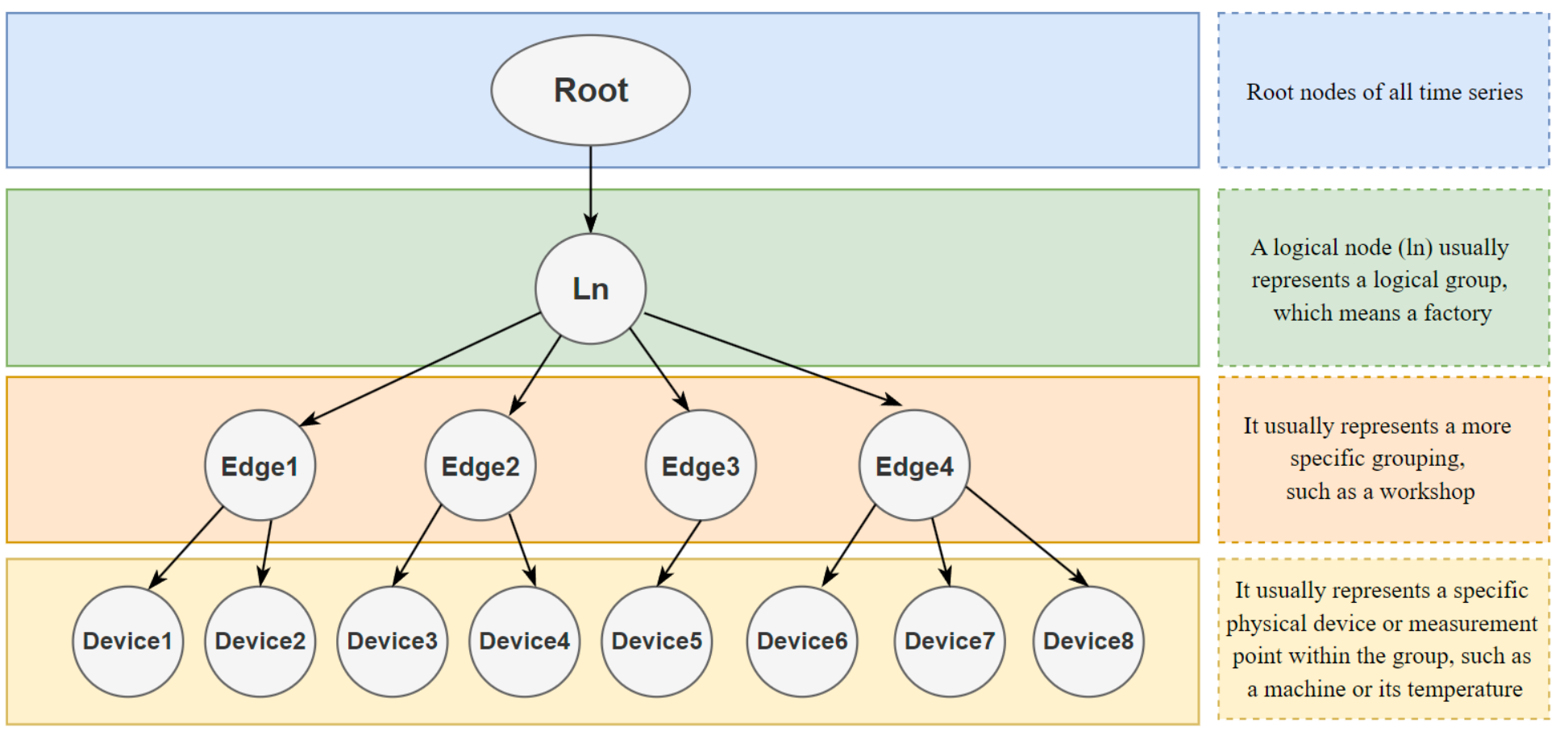}
    \caption{Explanation of the time series path}
    \label{fig:4-1}
\end{figure}

When the cloud receives a task migration request, the primary challenge is to verify whether the target query data is accessible in the cloud cache. To address this, we adopt a global localization mechanism based on hierarchical data partition rules. Specifically, each \textit{DataNode} serves as a parent node, and the time series segments that it manages are treated as leaf nodes to construct a hierarchical path tree (e.g., root.ln.edge1.device1). The explanation of the time series path is shown in Figure~\ref{fig:4-1}.

By parsing the time series path, the system determines the physical location of the data. The cloud maintains a metadata mapping table of $\langle \text{path} \rightarrow \text{edge \textit{DataNode}} \rangle$, and upon receiving a migration request, it performs path matching to validate cache hits.

\subsection{real-time consistency maintenance}\label{subsec4.2}

\begin{figure}[h!]
    \setlength{\belowcaptionskip}{-15pt}
    \centering
    \includegraphics[scale=0.3]{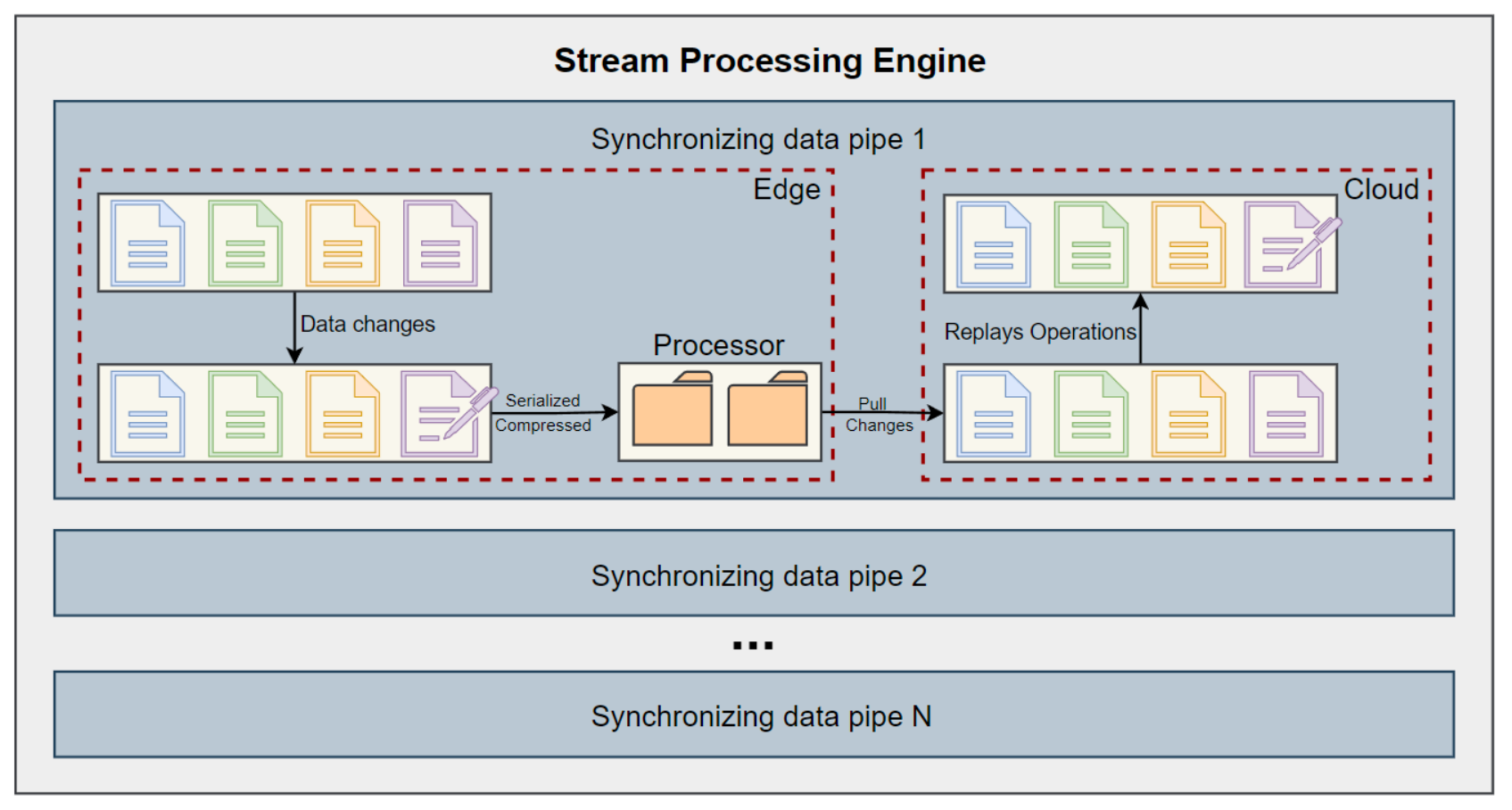}
    \caption{Workflow of the asynchronous delta streaming pipe}
    \label{fig:4-2}
\end{figure}

To ensure consistency between cloud and edge data upon cache hit, we adopt an asynchronous delta streaming pipe~\cite{b14}. As shown in Figure~\ref{fig:4-2}, a stream processing engine continuously captures data changes (inserts, deletes, updates) from the edge \textit{DataNode} in real time. These change records are serialized and batch-compressed by the processor, then pushed to a distributed message queue. The cloud cache engine subscribes to the queue and replays the operations in transaction order.
This pipe guarantees strong eventual consistency between the edge \textit{DataNode} and the cloud cache while maintaining low synchronization latency.

\subsection{collaborative dynamic data indexing}\label{subsec4.3}

\begin{figure}[h!]
    \setlength{\belowcaptionskip}{-20pt}
    \centering
    \includegraphics[scale=0.32]{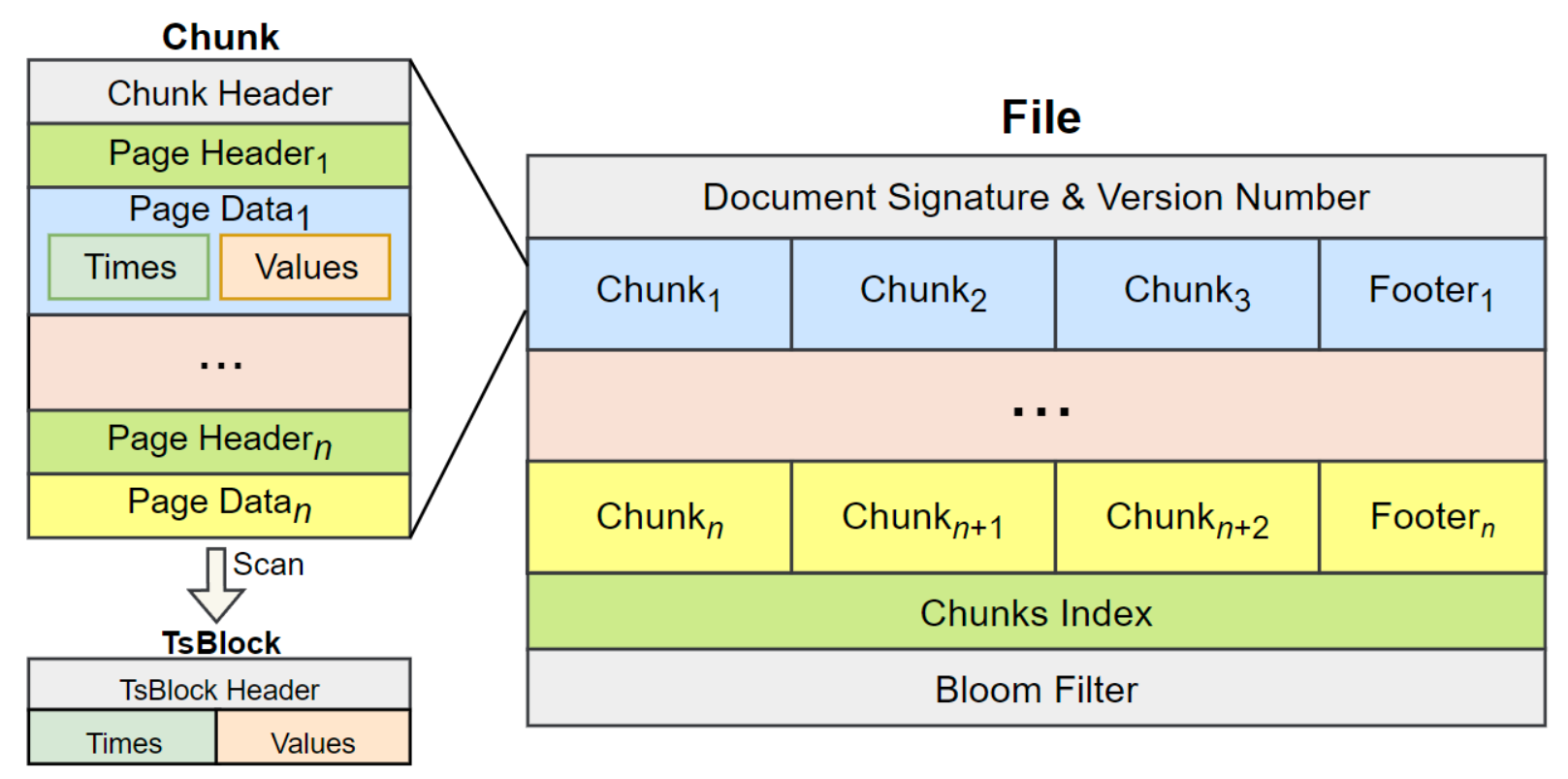}
    \caption{Structure of the File, Chunk and Page}
    \label{fig:4-3}
\end{figure}

To precisely match the execution progress after task migration and eliminate redundant queries, we introduce a logical indexing mechanism tailored to data access patterns. In the hierarchical storage structure, physical data units are organized in increasing granularity as follows: \textit{Page} (composed of \textit{TsBlocks}), \textit{Chunk} (aggregating multiple \textit{Pages}), and \textit{File} (containing multiple \textit{Chunks}). Data reads are orchestrated by the \textit{TsFile} component in collaboration with the scan operator. At scan operator startup, \textit{TsFile} sequentially loads the first \textit{Page} of the first \textit{Chunk} of the first \textit{File}, and iterates accordingly. \textit{Chunk} loading triggers the prefetching of all its \textit{Pages} into memory, while \textit{Page} access yields all \textit{TsBlocks} to the scan operator in bulk. The structure of the \textit{File}, \textit{Chunk} and \textit{Page} is shown in Figure \ref{fig:4-3}.

Based on this access pattern, we define the \textit{Chunk} as the minimum granularity unit for logical indexing. A \textit{Chunk} can only be indexed once all its \textit{Page}s have been fully consumed to avoid repeated access due to partial loading. However, directly transmitting \textit{Chunk} indices incurs significant overhead on the cloud, as it requires redundant metadata parsing. In line 5 of Algorithm~\ref{algo1}, to address this, we convert the \textit{Chunk} index into an equivalent data offset, which represents the cumulative number of processed bytes, and inject this offset as a filter predicate during query operator tree generation. This mechanism allows the cloud scan operator to skip all data preceding the given offset without reparsing the original \textit{Chunk} metadata, thereby achieving efficient and accurate resume-from-breakpoint execution with zero metadata parsing overhead.

\begin{algorithm}
\caption{Read TsBlcoks with the Chunk Offset}\label{algo1}
\KwIn{curOffset, Chunk}
\KwOut{The TsBlock pointed to by the offset}
\If{$Chunk \neq \text{null}$}{
    $queryFilter \gets getQueryFilter()$\;
    \If{$queryFilter = \text{null}$}{
        $rowCount \gets Chunk.getCount()$\;
        \If{$curOffset >= rowCount$}{
            $skipCurrentChunk()$\;
        }
        $curOffset \gets curOffset - rowCount$\;
    }
    \ElseIf{$\neg queryFilter.isSatisfied()$}{
        $skipCurrentChunk()$\;
    }
}
\end{algorithm}

\begin{figure}[h!]
    \setlength{\belowcaptionskip}{-20pt}
    \centering
    \includegraphics[scale=0.23]{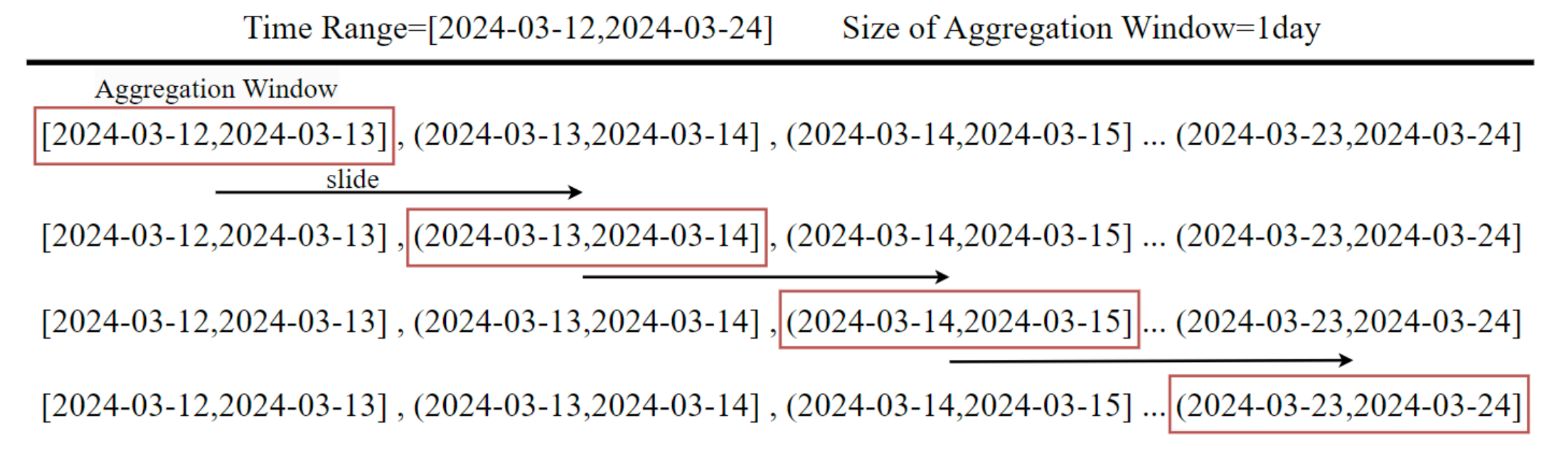}
    \caption{Workflow of aggregation scan operator}
    \label{fig:4-4}
\end{figure}

\begin{figure}[h!]
    \setlength{\belowcaptionskip}{-20pt}
    \centering
    \includegraphics[scale=0.32]{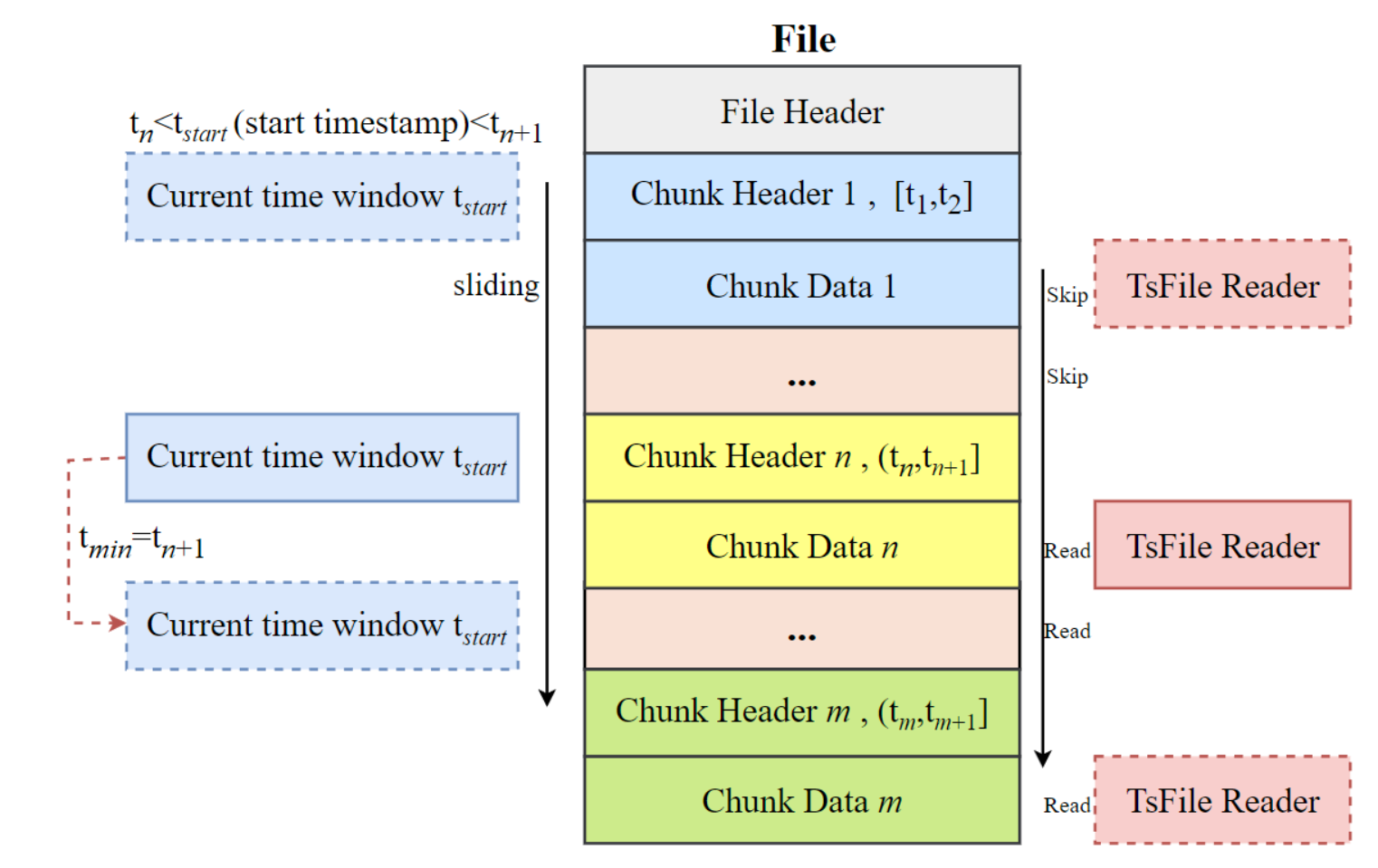}
    \caption{Workflow of sliding window}
    \label{fig:4-5}
\end{figure}

In addition to the series scan operator, we also design a collaborative aggregation scan operator. Its logical indexing mechanism must accommodate its time-window-driven execution semantics. As shown in Figure \ref{fig:4-4}, the aggregation scan operator partitions the query time domain into equal-length windows via a time range iterator and performs windowed aggregation (e.g., MAX/COUNT) in a sliding window fashion. Unlike the series scan operator, the aggregation scan operator conducts a pre-check on each data block's timestamp range before execution. As shown in Figure \ref{fig:4-5}, if a block lies entirely before the current window, metadata-level skipping is triggered to avoid unnecessary I/O. Only blocks overlapping the window are loaded and processed. Consequently, we define the start timestamp of the current window as the logical index. This design ensures that data earlier than the index is excluded from processing, while preserving the continuity of windowed computations.

By precisely defining the logical index, we enable lightweight packaging of query deltas. These deltas allow the cloud to accurately resume execution from the edge breakpoint, ensuring that only unconsumed data units are processed after migration. This achieves dual objectives of zero state reconstruction overhead and zero redundant computation, thereby ensuring lossless continuity in cross-tier task migration.
\vspace{-10pt}

\section{Collaborative Scan Operator Architecture}\label{sec5}

To support dynamic task migration in the collaborative scan operator framework, the distributed scan operator must address the following challenges:(i) ensuring consistent query parsing and execution semantics across tiers; (ii) implementing an adaptive data transmission pipe that selects the appropriate mode based on query characteristics; and (iii) enabling pipe creation and isolation in multi-scan-operators scenarios.

\subsection{consistent query parsing and execution semantics}\label{subsec5.1}
To ensure query task consistency during dynamic task migration from edge to cloud, the execution plan generated on the cloud must be completely equivalent to the original query plan on the edge. We achieve this through a two-tier guarantee mechanism. First, based on the LRU synchronization protocol described in Section~\ref{subsec4.2}, the cloud cache and the edge hotspot \textit{DataNode} maintain strong data consistency. Second, since both cloud and edge \textit{DataNodes} act as metadata holders~\cite{b39} and share a globally unified data directory composed of the full set of storage group metadata. By this point, the accessibility of target data in the cloud has already been validated by hierarchical path mapping in Section~\ref{subsec4.1}, and the metadata view is fully synchronized. Therefore, the cloud can directly parse the original SQL statement to generate an equivalent query plan at migration time. This process preserves semantic consistency across three critical dimensions, operator selection, join order, and data access paths, so that only the SQL statement needs to be transferred for migration, completely eliminating any risk of execution divergence across tiers.

\subsection{collaborative dynamic data indexing}\label{subsec5.2}
To enable precise and efficient transfer of query results between cloud and edge operators, we design an on-demand data channel mechanism. This channel is initialized when the resource monitor sends a migration instruction to the edge \textit{DataNode} and is activated upon receiving the cloud’s migration confirmation. When the edge scan operator receives the cloud’s migration instruction, which includes the cloud operator’s network endpoint (IP address and port), data fragment identifier (\textit{FragmentID}), and operator instance identifier (\textit{SourceID}), it establishes a low-latency point-to-point connection with the cloud. To maximize network bandwidth utilization, data is transmitted in \textit{TsBlocks} as the atomic unit, tightly integrated with Volcano model semantics: on the cloud \textit{Source}\footnote{Source is the point for sending data.}, each invocation of \textit{next()} produces a result \textit{TsBlock} that is immediately pushed into the transmit queue; on the edge \textit{Sink}\footnote{Sink is the point for pulling data.}, invoking \textit{next()} asynchronously pulls \textit{TsBlocks} from the receive queue, while \textit{hasNext()} continuously monitors the channel’s status flag. When the cloud finishes execution, it sends a termination marker to trigger channel closure; the edge operator’s \textit{hasNext()} detects this marker and releases resources immediately. For task rollback scenarios, the edge scan operator uses persisted context state and logical index (e.g., data offset or window start timestamp) to accurately resume data reading from the local storage, thereby avoiding redundant I/O operations.

\begin{figure}[h!]
    \setlength{\belowcaptionskip}{-20pt}
    \centering
    \includegraphics[scale=0.38]{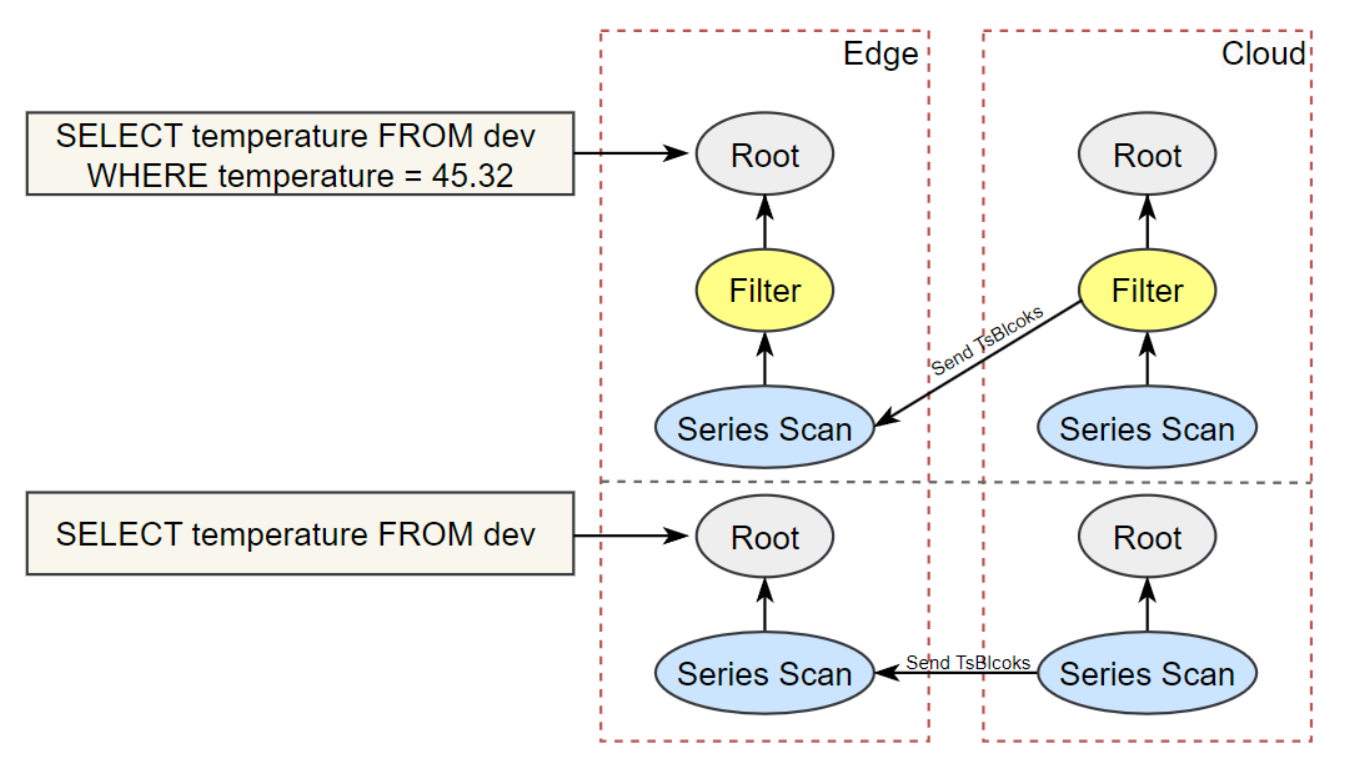}
    \caption{Predicate push down mode selection}
    \label{fig:5-1}
\end{figure}

However, in practice, network bandwidth constraints become the critical bottleneck for cloud-edge collaboration. To alleviate transmission pressure, we design an adaptive multi-modal transmission mechanism. For large-scale scan queries without filter predicates, we employ the block-streaming mode described in Section~\ref{subsec3.2}, returning results in batches of 1,000 records per \textit{TsBlock}. The client requests subsequent batches on demand, decomposing massive transfers into micro-batches to reduce per-transfer load. However, this mode fails for queries containing filter predicates (e.g., WHERE clauses). If filtering were performed at the edge, the cloud would need to transfer entire \textit{TsBlocks}, wasting bandwidth. To address this issue, we introduce a push-down optimization that executes filter operations on the cloud, sending only predicate-satisfying data back to the edge in \textit{TsBlocks}. As shown in Figure~\ref{fig:5-1}, when a query plan includes a filter operator, the system automatically switches to the predicate pushdown mode. By dynamically identifying query characteristics and switching transmission strategies, we achieve a truly network-aware, adaptive pipe mechanism.

\subsection{Collaborative dynamic data indexing}\label{subsec5.3}

When processing query plans involving multiple scan operators (e.g., SELECT humidity, temperature FROM root.ln.edge1.device1, which requires parallel execution of two scan operators), pipe isolation becomes a key challenge for ensuring query correctness. Based on the channel framework established in Section~\ref{subsec5.2}, cloud-edge data transmission connections are identified using a four-element tuple $\langle \text{IP},\text{Port},\text{\textit{FragmentID}},\text{\textit{SourceID}} \rangle$. Since IP and port are constant within a single query, the \textit{FragmentID} and \textit{SourceID} serve as the core dimensions for operator-level isolation.

Each operator instance registers its $\langle \text{\textit{FragmentID}},\text{\textit{SourceID}} \rangle$ pair in a global mapping table during initialization. The cloud then uses this table to establish an accurate mapping $\langle \text{CloudIdentifier} \rightarrow \text{EdgeIdentifier} \rangle$, enabling precise linkage between cloud and edge operator instances. However, under concurrent multi-query, this mapping alone is insufficient to disambiguate scan operators across different queries.

To address this, the system introduces a globally unique query identifier (\textit{QueryID}) and extends the connection identifier to a five-element tuple $\langle \text{IP},\text{Port},\text{\textit{FragmentID}},\text{\textit{SourceID}},\text{\textit{QueryID}}\rangle$. This enables three-level orthogonal isolation:\textit{QueryID} partitions the boundaries of different query tasks; \textit{FragmentID} + \textit{SourceID} isolates parallel operators within the same query; and IP + port restricts connections to a specific physical node resource domain. This design ensures accurate and fine-grained pipe governance for complex multi-operator scenarios.

\section{Delta State Synchronization Protocol}\label{sec6}

To enable seamless operator migration between cloud and edge, a lightweight incremental state synchronization mechanism is required. This mechanism comprises: (i) real-time exchange of lightweight execution progress; (ii) an asynchronous compensation strategy to tolerate network variability; and (iii) timely consistency guarantees for delta state updates.

\subsection{real-time exchange of lightweight execution progress}\label{subsec6.1}

To achieve real-time synchronization of lightweight execution progress between cloud and edge, an efficient transmission protocol and support for remote service invocation are essential. We adopt Thrift~\cite{b33} as the core communication framework. As a cross-platform Remote Procedure Call (RPC) protocol~\cite{b34}, Thrift enables interaction at the function level between heterogeneous systems through a client-server architecture. The server listens on a designated port and registers callable functions, while the client initiates remote invocations and receives return values. This protocol provides foundational support for state synchronization during operator migration.

\begin{figure}[h!]
    \setlength{\belowcaptionskip}{-10pt}
    \centering
    \includegraphics[scale=0.38]{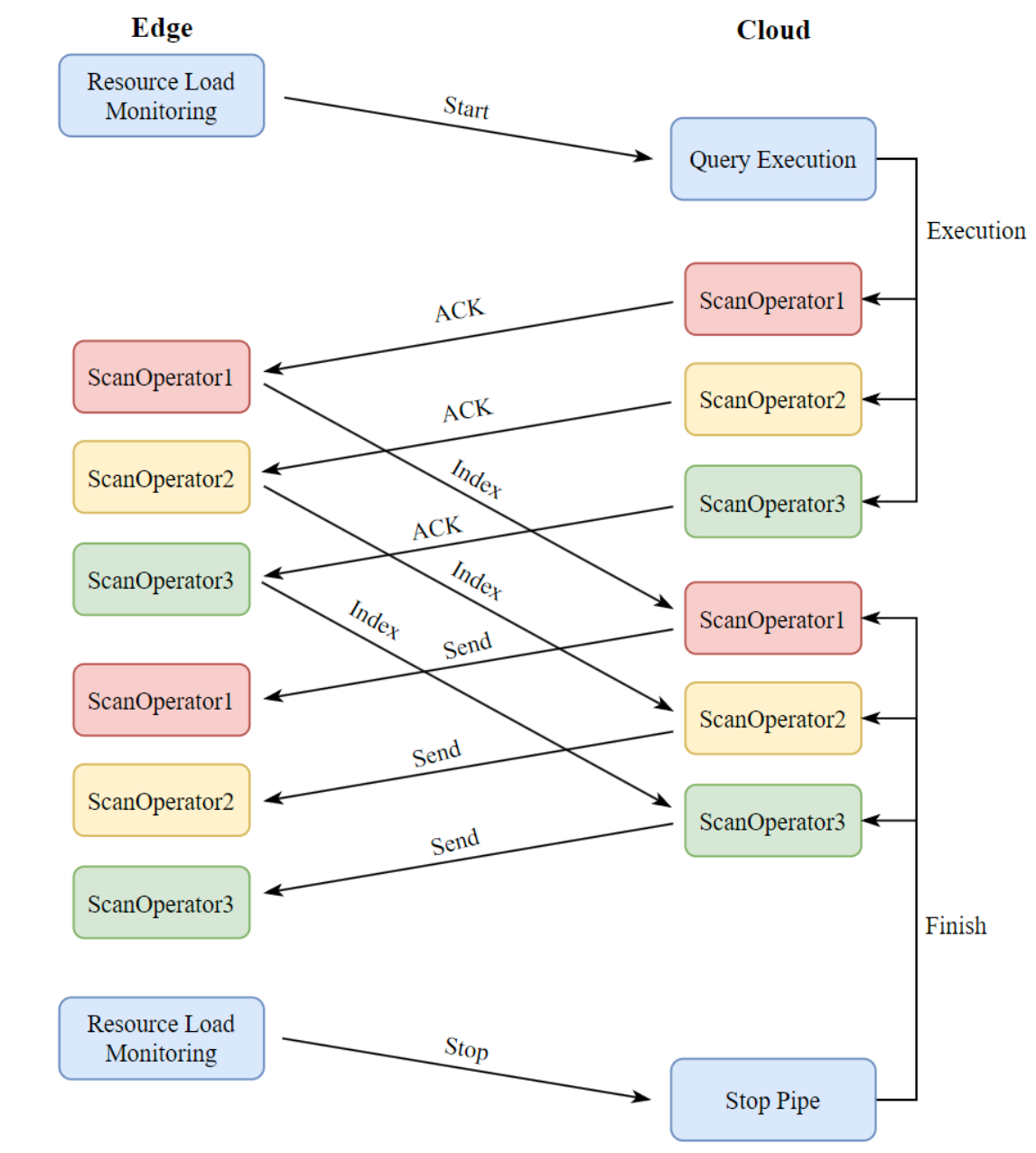}
    \caption{Mechanism for real-time exchange}
    \label{fig:6-2}
\end{figure}
As shown in Figure~\ref{fig:6-2}, the mechanism for real-time exchange of lightweight execution progress during task migration is as follows:

\begin{enumerate}
    \item Upon receiving a migration decision from the resource monitor, the edge \textit{DataNode} sends a migration request to the cloud, which includes a quintuple identifier $\langle \text{IP}, \text{Port}, \text{FragmentID}, \text{SourceID}, \text{QueryID} \rangle$ along with the original SQL statement.
    \item After receiving the request, the cloud immediately constructs the corresponding query plan and instantiates the scan operator. A confirmation message is then returned to the edge, containing a triplet $\langle \text{FragmentID}, \text{SourceID}, \text{QueryID} \rangle$ (the cloud IP/port is omitted as it is fixed).
    \item Upon receiving confirmation, the edge stops reading from local data and serializes the current delta execution state, which is sent to the cloud.
    \item The cloud initializes the operator context using the delta state and resumes the query from the breakpoint, returning results in a \textit{TsBlock} streaming manner.
    \item The edge operator pulls data blocks on demand through the \textit{next()} interface, achieving a transparent switch in execution.
    \item When the resource monitor issues a remigration command or the cloud completes the query, a termination flag is sent to close the transmission pipe.
\end{enumerate}

This design enables millisecond-level synchronization latency, ensuring real-time exchange of lightweight execution progress.

\subsection{asynchronous compensation strategy}\label{subsec6.2}

To address execution efficiency challenges under dynamic network conditions, we design a fault-tolerant compensation mechanism based on asynchronous waiting. After the edge initiates a task migration instruction, the edge \textit{DataNode} continues local data reading until a readiness confirmation signal is received from the cloud, at which point it switches to the cloud data stream. This approach prevents idle edge resources caused by cloud query plan generation delays.

To ensure data transmission integrity, the pipe also adopts a TCP-like handshake confirmation protocol. Once the transmission pipe is established, the \textit{Sink} first sends a \textit{Future}, lightweight heartbeat probe packet, which is implemented as a header-only \textit{TsBlock}. The \textit{Source} asynchronously waits until it successfully receives the probe and returns an ACK, after which formal data transmission begins. This design significantly improves system robustness in highly volatile network environments.

\subsection{delta state updates timely consistency guarantees}\label{subsec6.3}

To accelerate collaborative update of query incremental states between the cloud and edge while ensuring strong consistency, we restructure the data reading workflow of the scan operator by introducing a real-time breakpoint tracking mechanism.

In series scan operators, after each \textit{Chunk} of data is loaded, the number of tuples read so far is recorded as a logical index. For aggregation scan operators, the start timestamp of the current sliding window serves as the index. To avoid redundant data reading and ensure that in-memory data is fully consumed, the edge \textit{DataNode} packages and sends the logical index as a delta state update only after the corresponding in-memory \textit{TsBlocks} have been completely consumed. During task remigration, the same principle applies that the data transmission pipe is closed only after the cloud \textit{TsBlocks} have been fully consumed. 

An event-driven asynchronous RPC invocation model is adopted, employing a non-blocking polling strategy. When the delta state is not yet updated, the request thread is suspended, and a callback function is registered to immediately trigger subsequent operations once an update is detected. These designs ensure real-time and consistent delta state updates, maintaining correctness throughout the operator migration process.

\section{Experiments}\label{sec7}
In this section, we compare the CED collaborative scan operators (series scan operator and aggregation scan operator) with the native IoTDB scan operators, including: (1) query latency under different I/O loads; (2) query latency under high concurrency; (3) query latency under different network bandwidths. All the encoding methods are implemented in Cloud-Edge-Device DataBase (CED-DB)\footnote{https://github.com/MDC-LOADS/Cloud-Edge-Device-DataBase}.

\subsection{Experimental Settings}\label{subsec7.1}

\subsubsection{Datasets}\label{subsubsec7.1.1}

The datasets are generated by the IoTDB-benchmark~\cite{b35}, a time series database benchmark specialized in generating IoT workloads, based on data from the train application. The total number of sensors is nine. These sensors include boolean variables, strings, integer variables, and floating-point numbers, which change randomly over time. All data strictly follow an increasing timestamp order, ensuring that there are no out-of-order data occurrences, with a sampling interval of 0.001 seconds. The total generated data size is 3GB.

\subsubsection{Baselines and Evaluation Metrics}\label{subsubsec7.1.2}

In this paper, the performance evaluation of the query is benchmarked against the native query performance of the Apache IoTDB-1.3.0 database. The evaluation metrics are defined as follows:

\begin{itemize}
    \item \textbf{I/O Utilization Percentage (I/O Usage):} This metric measures the I/O utilization based on the read rate compared to the maximum read rate. The calculation formula used is:

    \begin{equation}
        \text{\textit{I/O Usage}} = \frac{\text{\textit{Read Rate}}}{\text{\textit{Max Read Rate}}}
    \end{equation}

    \textit{Read Rate} refers to the current rate of data reads during queries, and \textit{Max Read Rate} is the maximum achievable read rate under optimal conditions.

    \item \textbf{Query Performance}~\cite{b36}\textbf{ (Query Execution Time):} This metric evaluates the average query execution time per query. The calculation formula used is:
    \begin{equation}
        \text{\textit{Query Execution Time}} = \frac{\text{\textit{Total Time}}}{\text{\textit{Total Number of Queries}}}
    \end{equation}

    \textit{Total Time} represents the cumulative time taken to execute all queries, and \textit{Total Number of Queries} is the count of queries executed during the evaluation period.
    \item \textbf{Queries Per Second (QPS):} This metric evaluates the average number of queries per second. The calculation formula used is:
    \begin{equation}
    \text{\textit{QPS}} = \frac{\text{\textit{Number of Parallel Queries}}}{\text{\textit{Max Time}}}
    \end{equation}
    \textit{Max Time} refers to the execution time of the longest query in parallel execution, and \textit{Number of Parallel Queries} is the count of parallel queries executed during the evaluation period.
\end{itemize}

These metrics are used to compare and assess the query performance of the experimental setup against the baseline provided by Apache IoTDB-1.3.0. The aim is to measure how efficiently the CED-DB handles queries relative to the established database version.

\subsubsection{Experimental Setup}\label{subsubsec7.1.3}

We use Huawei Cloud Elastic Cloud Servers (ECS) to simulate a distributed query environment involving both cloud and edge nodes. A total of three servers are deployed: one as the cloud and two as edges.

The cloud server is equipped with a 16-core CPU and 32 GB of memory, simulating a powerful centralized computing environment. Each edge server is configured with a 4-core CPU, 16 GB of memory, and dual 2 TB HDDs, emulating resource-constrained edge computers. All three servers are connected within the same local area network (LAN).

In terms of system deployment, the cloud server hosts an independent cluster consisting of one \textit{ConfigNode} and one \textit{DataNode}. The two edge servers form a separate edge cluster, comprising one shared \textit{ConfigNode} and one \textit{DataNode} on each server. This setup establishes a basic cloud–edge collaborative architecture for our experiments. The machine configurations are detailed in Table~\ref{tab:7-1}.

\begin{table}[htbp]
\centering
\caption{Hardware Configuration of Cloud and Edge Servers}
\label{tab:7-1}
\begin{tabular}{lllll}
\toprule
\textbf{Server Type} & \textbf{CPU} & \textbf{Memory} & \textbf{Storage} & \textbf{Quantity} \\
\midrule
Cloud Server & 16 cores & 32 GB & 40 GB SSD & 1 \\
Edge Server & 4 cores & 16 GB & 2 × 4 TB HDD & 2 \\
\bottomrule
\end{tabular}
\end{table}

\subsection{Performance Evaluation}\label{subsec7.2}

To evaluate the performance advantages of the CED collaborative scan operator, we compare three execution modes: pure cloud execution, pure edge execution, and CED collaborative execution. To cover a range of query workload characteristics, five representative query types are selected: (1) single-field string filter scan; (2) single-field float filter scan; (3) multi-field large-scale scan; (4) string-based time-window aggregation; and (5) float-based time-window aggregation. The detailed query SQLs are listed in Table~\ref{tab:7-2}. All experiments are conducted in an environment with unrestricted network bandwidth, and average query execution time is used as the primary evaluation metric. 
\begin{table}[htbp]
\centering
\caption{Five Representative Query Types}
\label{tab:7-2}
\begin{tabular}{ll}
\toprule
Query Name & SQL \\
\midrule
Q1 & \texttt{SELECT t1 FROM dev WHERE t1='v999'} \\
Q2 & \texttt{SELECT t3 FROM dev WHERE t3=497.44467} \\
Q3 & \texttt{SELECT t1, t3 FROM dev} \\
Q4 & \texttt{SELECT count(t1) FROM dev} \\
                                    & \texttt{GROUP BY 5m} \\
Q5 & \texttt{SELECT max\_value(t3) FROM dev} \\
                                    & \texttt{GROUP BY 5m} \\
\bottomrule
\end{tabular}
\end{table}
\vspace{-10pt}
\begin{figure}[h!]
    \setlength{\belowcaptionskip}{-8pt}
    \centering
    \includegraphics[scale=0.33]{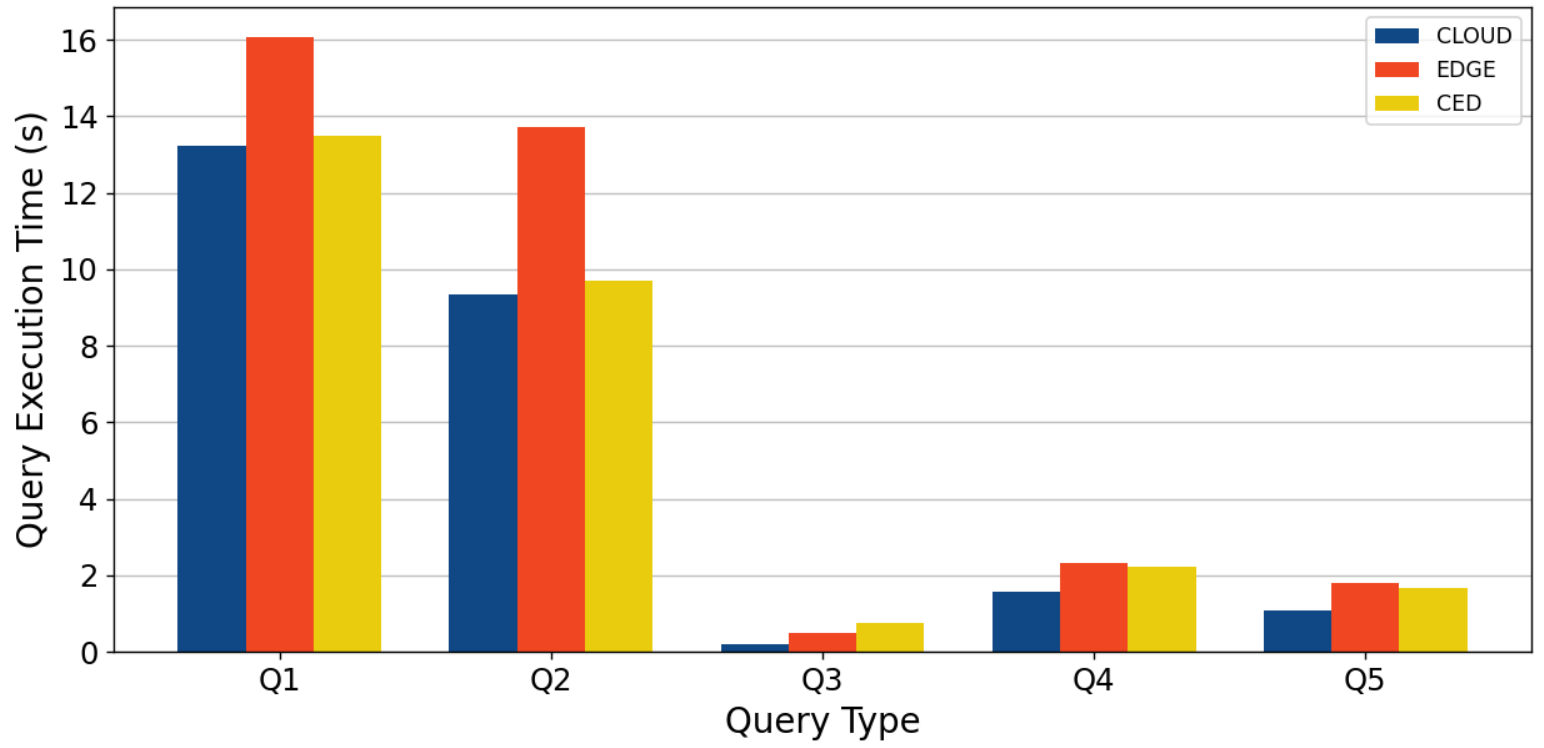}
    \caption{Comparison of query execution time under different query types}
    \label{fig:7-1}
\end{figure}

As shown in Figure~\ref{fig:7-1}, (a) Cloud Execution Advantage: Benefiting from high-performance server resources, cloud-only execution consistently yields lower latency than edge-only execution, as expected; (2) Divergent Effects of Collaborative Execution: For queries with small result sets (Q1/Q2/Q4/Q5), collaborative execution reduces latency by 19.7\% to 34.2\% compared to edge-only execution. This is primarily due to the cloud's computational advantage outweighing the network overhead. However, for large-scale multi-field scans (Q3), collaborative execution increases latency by 33.1\%. This is because the data transmission volume scales with the number of scan operators (as per Section 3.2, each operator transmits 1,000 rows per request), making network transfer the dominant bottleneck; (3) Overall Efficiency: Across all five query types, the collaborative execution mode achieves a weighted average latency reduction of 7.6\%, demonstrating its general effectiveness and applicability in scan-intensive workloads.

\subsection{High I/O Load Performance Evaluation}\label{subsec7.3}

\begin{figure}[h!]
    \setlength{\belowcaptionskip}{-16pt}
    \centering
    \includegraphics[scale=0.33]{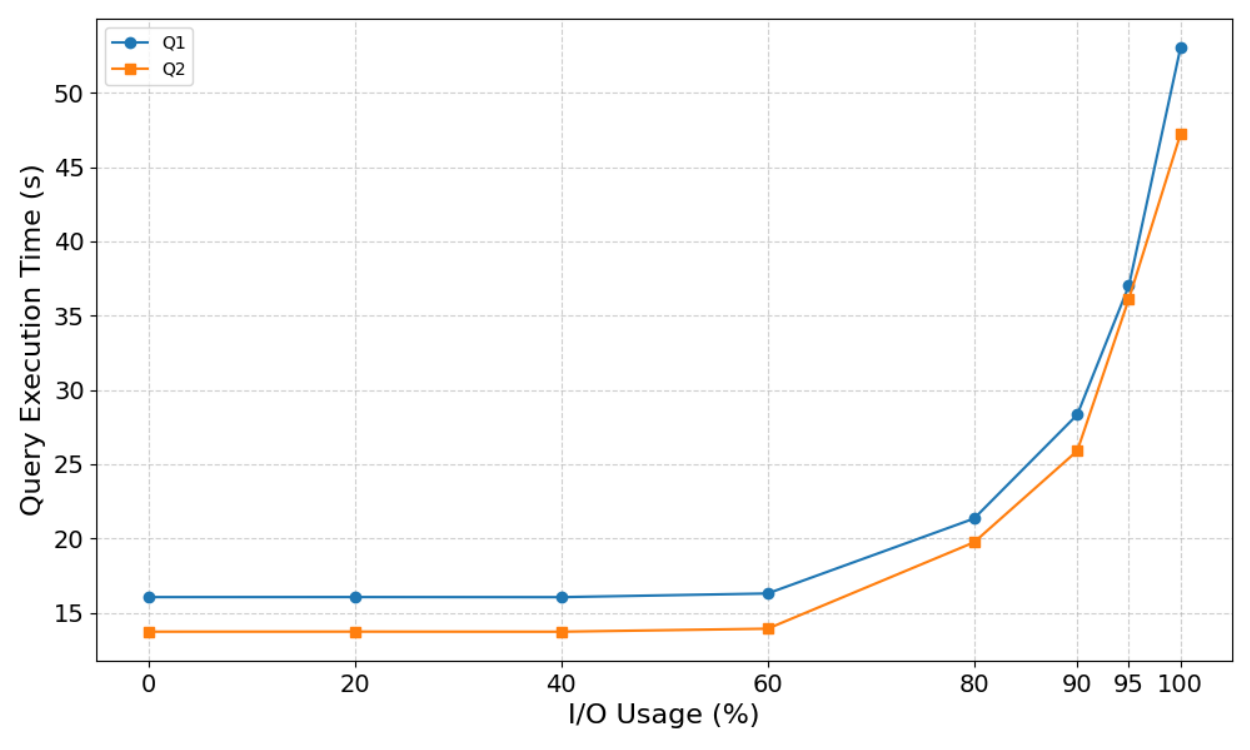}
    \caption{Query execution time under different I/O Load (Q1, Q2)}
    \label{fig:7-2}
\end{figure}

\begin{figure}[h!]
    \setlength{\belowcaptionskip}{-10pt}
    \centering
    \includegraphics[scale=0.33]{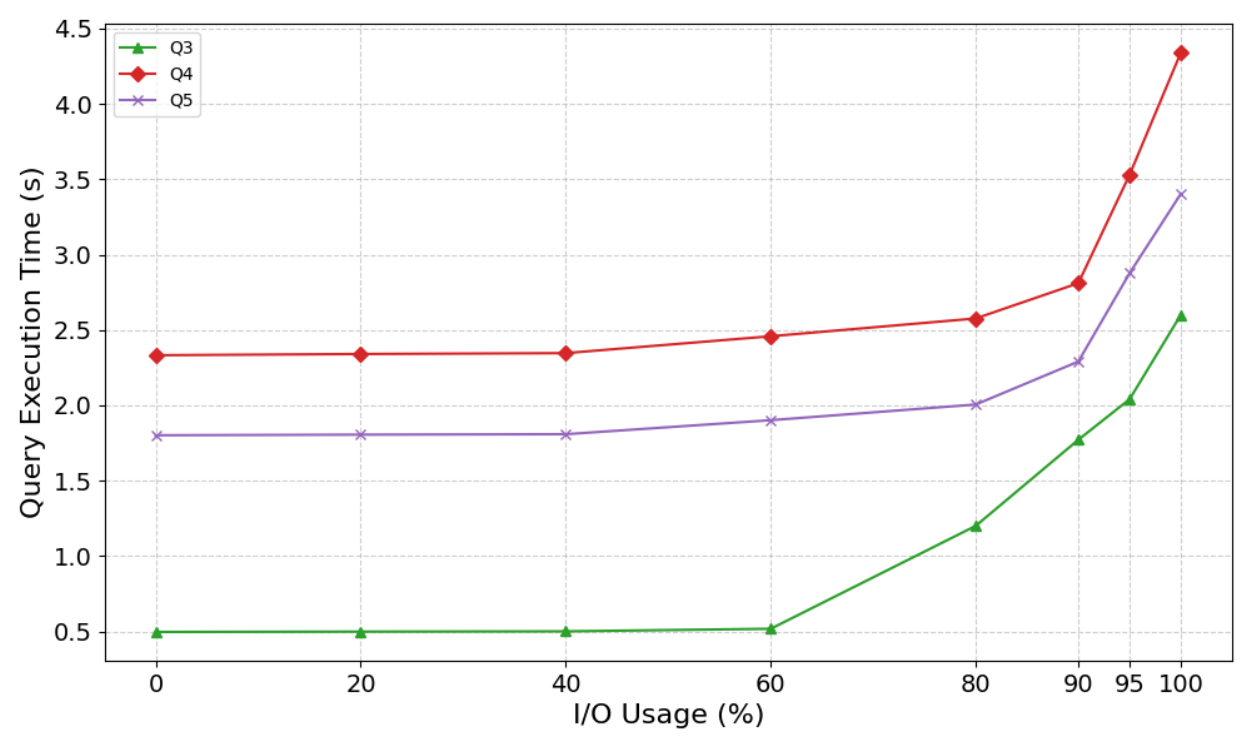}
    \caption{Query execution time under different I/O Load (Q3,Q4,Q5)}
    \label{fig:7-3}
\end{figure}

To evaluate the impact of high I/O load on edge-side query performance, we conduct experiments in which the I/O utilization of the edge is gradually increased. Under five standard query workloads, we compare pure edge execution with CED collaborative execution. As shown in Figure~\ref{fig:7-2} and \ref{fig:7-3}, when the I/O utilization exceeds the 80\% threshold, the collaborative mode demonstrates significant performance advantages across all query types.

Specifically, pure edge execution suffers from notable latency increases due to storage bottlenecks, with an average query latency rise of 25.2\% and a maximum increase of 43.7\% observed in Q2. In contrast, the CED collaborative execution mode reduces query latency by an average of 32.5\%, achieving a maximum reduction of 49.2\% in Q2 by offloading query tasks to the cloud.

These results validate the core value of our collaborative operator design: under high I/O usage (above 80\%), the dynamic task migration mechanism effectively bypasses storage bottlenecks, boosting query efficiency to 1.63× that of the edge-only baseline. Overall, the experimental results confirm the strong adaptability of the CED collaborative architecture in I/O-constrained edge environments.

\subsection{High CPU Load Performance Evaluation}\label{subsec7.4}
\begin{figure}[h!]
    \setlength{\belowcaptionskip}{-14pt}
    \centering
    \includegraphics[scale=0.33]{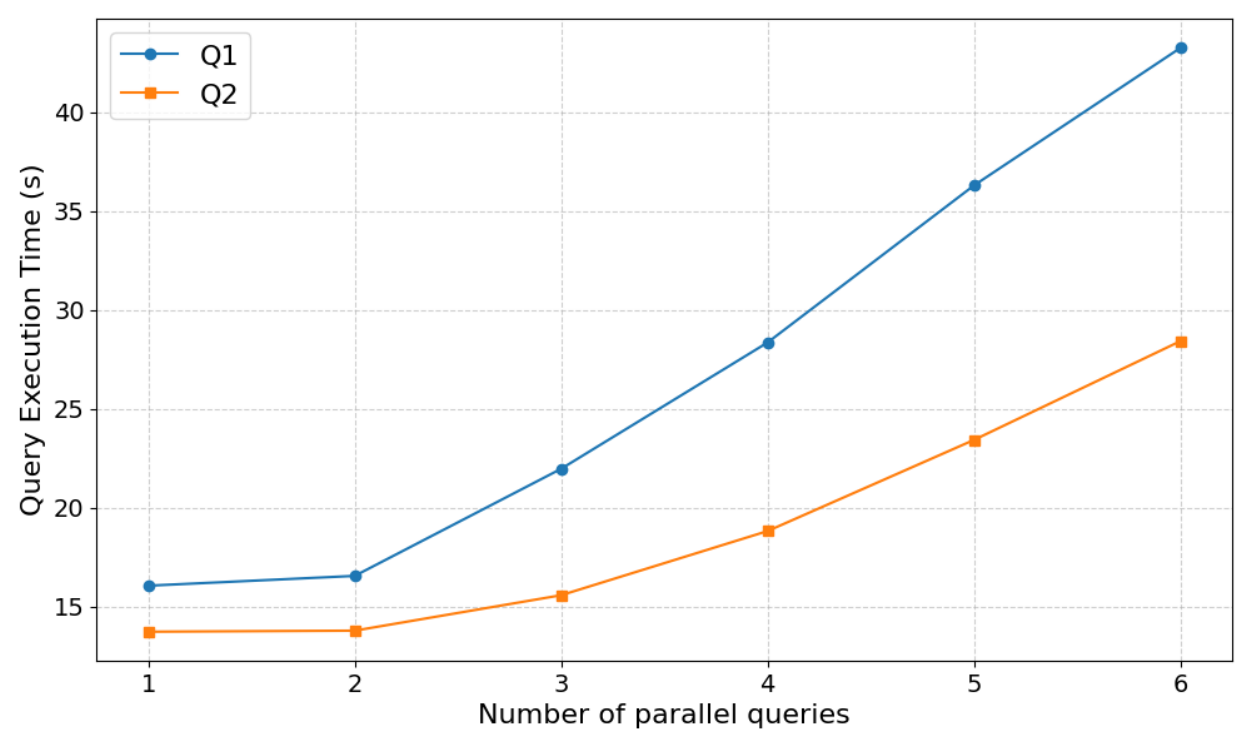}
    \caption{Query execution time under different CPU Load (Q1, Q2)}
    \label{fig:7-4}
\end{figure}

\begin{figure}[h!]
    \setlength{\belowcaptionskip}{-10pt}
    \centering
    \includegraphics[scale=0.33]{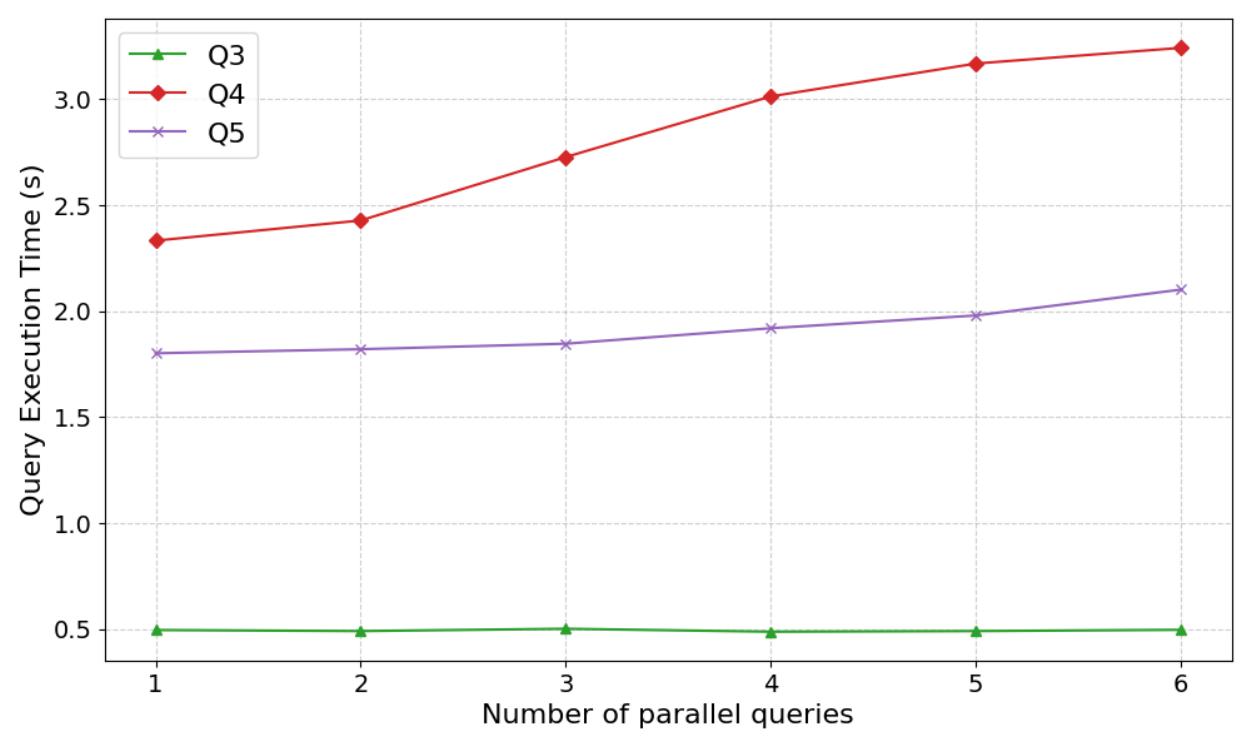}
    \caption{Query execution time under different CPU Load (Q3,Q4,Q5)}
    \label{fig:7-5}
\end{figure}

To evaluate the impact of CPU load on edge query performance, we gradually increase the number of concurrent queries (from 1 to 6) to drive up CPU utilization. Under five representative query workloads, we compare pure edge execution with the CED collaborative mode. As shown in Figure \ref{fig:7-4} and \ref{fig:7-5}, the edge query latency increases nonlinearly with growing concurrency. This trend is primarily driven by scheduling delays induced by CPU resource contention—particularly for compute-intensive queries such as string filters(Q1). At 4-way concurrency, the latency of Q1 and Q2 increases by 76.6\%. Aggregation queries (Q4 and Q5) exhibit a smaller latency increase of 29.1\% due to moderate overhead from window computations. In contrast, large-scale full scans (Q3) show only a 1.2\% increase in latency under 6 concurrent queries, reflecting their low CPU sensitivity due to I/O-bound characteristics.

In this scenario, the CED collaborative mode demonstrates clear advantages. Taking Q1 at 4-way concurrency as an example, collaborative execution improves query efficiency by 92.4\% compared to edge-only execution. This benefit is attributed to the dynamic offloading of compute-intensive operators to the cloud. When CPU utilization exceeds 85\%, the collaborative mode reduces query latency by an average of 52.1\% across all five query types, effectively relieving the edge node from computational bottlenecks. These results confirm the general adaptability of our collaborative architecture in CPU-constrained environments.

\subsection{High Network Load Performance Evaluation}\label{subsec7.5}
\begin{figure}[h!]
    \setlength{\belowcaptionskip}{-10pt}
    \centering
    \includegraphics[scale=0.33]{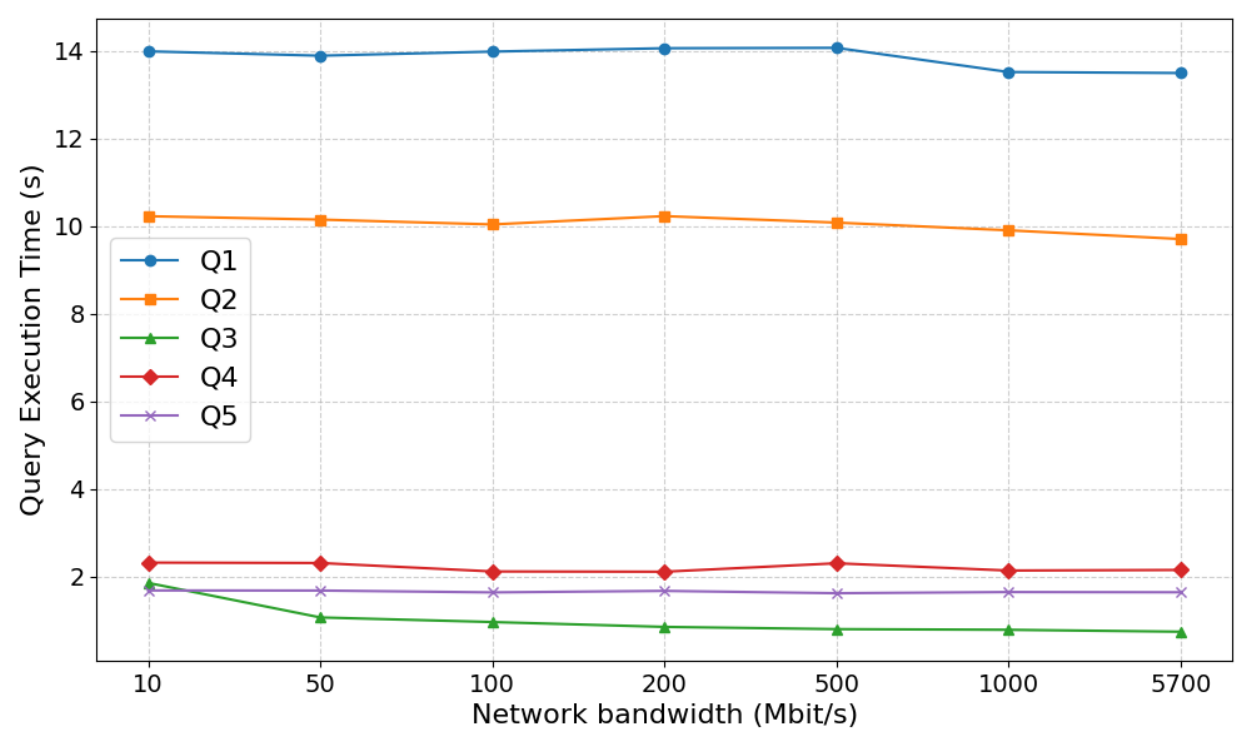}
    \caption{Query execution time under different Network Bandwidth}
    \label{fig:7-6}
\end{figure}

To evaluate the impact of constrained network bandwidth on the performance of CED collaboration, we systematically vary the available bandwidth under five standard query workloads and analyze the adaptability of the collaborative execution mode. As shown in Figure \ref{fig:7-6}, the effect of network limitations on performance exhibits strong query-type dependency.

For queries with small result sets (e.g., Q1/Q2, result size $<$ 100 records), reducing the bandwidth to 50 Mbps causes only a modest average latency increase of 4.2\%. Aggregation queries (Q4/Q5, ~600 result records) experience a 6.5\% latency increase under the same bandwidth, mainly due to the cost of result transmission. In contrast, large-scale scan queries (Q3, result size of 2×1000 records) show a significant 43.0\% increase in latency, clearly indicating sensitivity to data transmission volume.

In particular, even under the constrained 50 Mbps bandwidth, the collaborative execution mode still outperforms edge-only execution under high I/O load conditions ($>$80\% utilization), achieving a 12.4\% latency advantage. When bandwidth is restored to 500 Mbps, collaborative execution reduces latency by an average of 31.7\% across all query types except large-scale scans (Q3), compared to pure edge execution.

These findings confirm that under typical network conditions (bandwidth $>$ 50 Mbps), the CED collaborative architecture consistently outperforms edge-only execution in compute or storage constrained scenarios, demonstrating its broad applicability and performance benefits.

\subsection{Functionality and correctness Evaluation}\label{subsec7.6}

\subsubsection{Flexible Migration}\label{subsec7.6.1}

To evaluate the dynamic migration efficiency of the collaborative operator, we conduct an experiment where a string filter scan query (Q1) is initially executed at the edge, and a forced I/O overload (100\% utilization) is introduced at different time points to trigger migration. 
\vspace{-10pt}
\begin{figure}[h!]
    \setlength{\belowcaptionskip}{-10pt}
    \centering
    \includegraphics[scale=0.33]{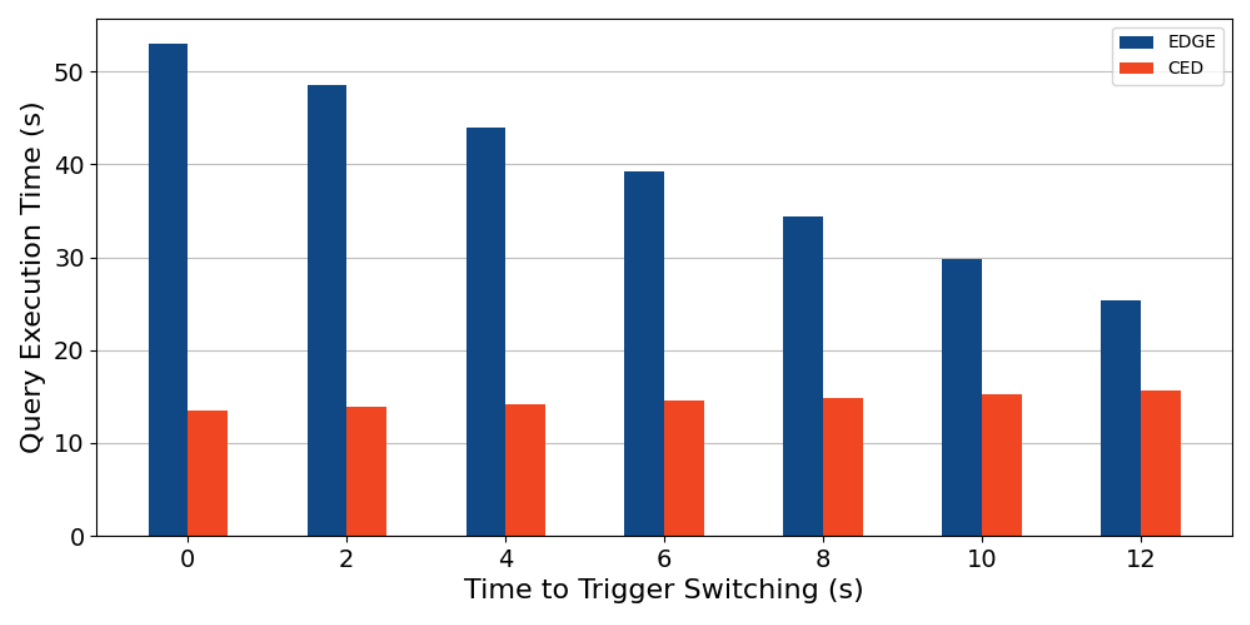}
    \caption{Query execution time under different time points}
    \label{fig:7-7}
\end{figure}

\begin{figure}[h!]
    \setlength{\belowcaptionskip}{-14pt}
    \centering
    \includegraphics[scale=0.33]{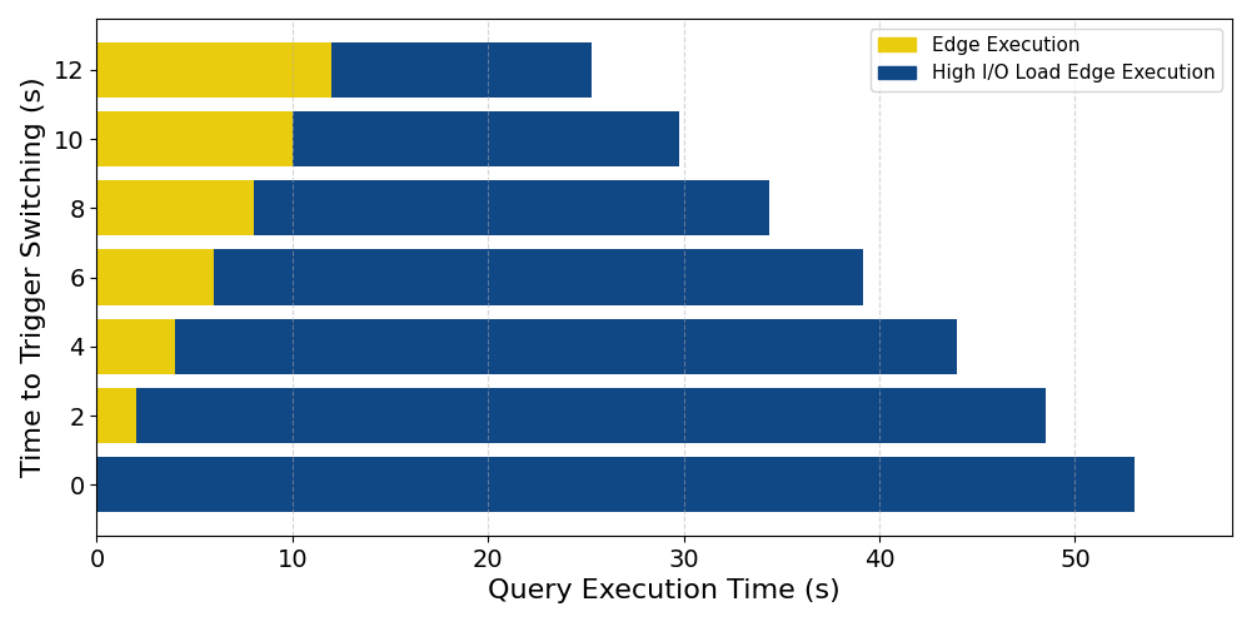}
    \caption{edge query execution time under different time points}
    \label{fig:7-8}
\end{figure}

\begin{figure}[h!]
    \setlength{\belowcaptionskip}{-14pt}
    \centering
    \includegraphics[scale=0.33]{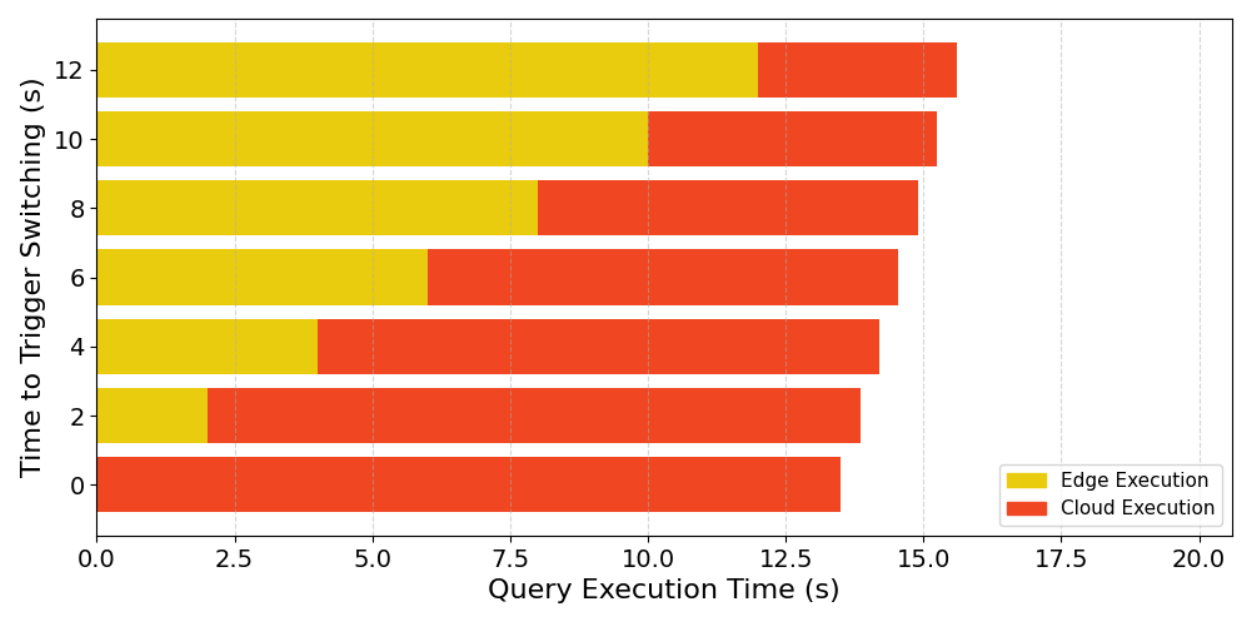}
    \caption{CED query execution time under different time points}
    \label{fig:7-9}
\end{figure}

As shown in Figure \ref{fig:7-7}, the query task is seamlessly migrated to the cloud during execution. Figures \ref{fig:7-8} and \ref{fig:7-8} clearly show the CED migration process. Compared to the baseline approach that continues execution solely on the edge, the collaborative mode reduces the total query latency by 130.8\%. This demonstrates that under high-load conditions, the cloud-edge collaborative operator significantly improves query performance by flexibly offloading tasks in response to system pressure.

\subsubsection{Impact of Cache Hits}\label{subsec7.6.1}

To quantify the impact of cache hit rate on collaborative performance, we deploy eight parallel string filter queries (6-way Q1) and control the cache hit rate by preloading hot data blocks into designated \textit{DataNodes}. 
\vspace{-10pt}
\begin{figure}[h!]
    \setlength{\belowcaptionskip}{-14pt}
    \centering
    \includegraphics[scale=0.45]{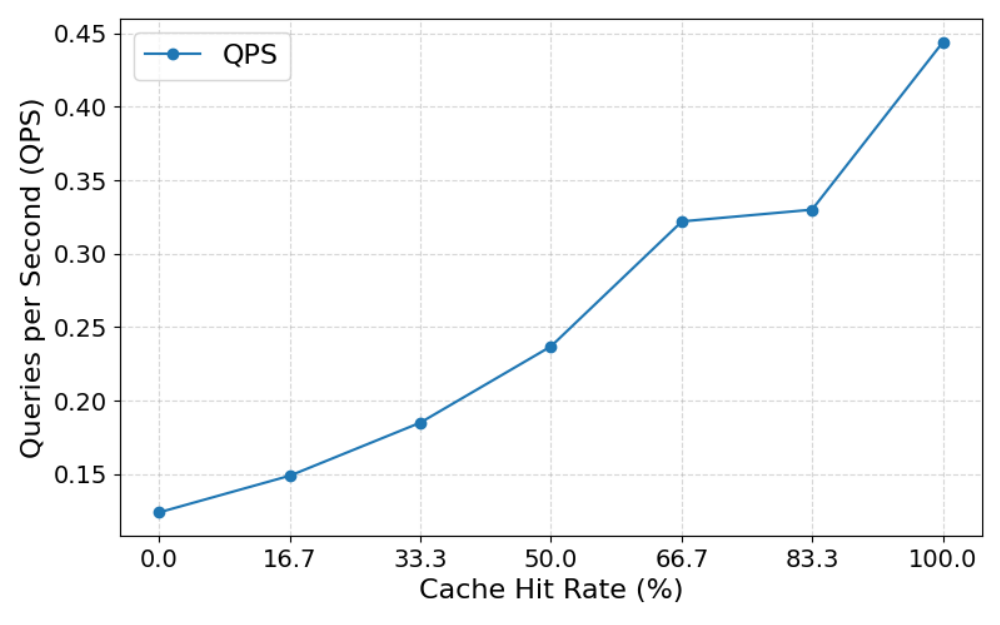}
    \caption{QPS under different Cache Hit Rate}
    \label{fig:7-10}
\end{figure}

As shown in Figure~\ref{fig:7-10}, the QPS increases steadily with higher cache hit rates, indicating reduced execution time and improved parallel query efficiency. This confirms that the cache hit rate is a critical amplification factor for collaborative performance. Specifically, when the cache hit rate reaches 50\%, the QPS increases by 91.1\%. Even with a low cache hit rate of 17\%, QPS still improves by 20.1\%, demonstrating that the CED collaborative operator maintains a significant advantage under large-scale data workloads.

\subsection{Experimental Summary}\label{subsec7.6}

In this experiment, we analyzed the impact of different factors such as I/O load, CPU load, network bandwidth at the edge. We compared the average query execution times under three scenarios: cloud, edge, and collaborative scan operator. The results show that under high I/O usage, the collaborative scan operator can improve query efficiency by an average of 22.6\%; under high CPU load, it can improve query efficiency by an average of 16.3\%; and under low network bandwidth, its efficiency decreases by 11.1\%. The integration of the collaborative scan operator fully utilizes both the cloud and edge, reducing their respective loads. This indicates that under conditions of good network download bandwidth, the collaborative scan operator can effectively mitigate the issue of slow scan operator reads or execution caused by high resource load. It can efficiently balance the resource scheduling between the cloud and edge.

\section{Conclusion}\label{sec8}

This paper rearchitects the query execution model of time-series databases by proposing a cloud-edge-device collaborative query framework and collaborative operator design, which significantly enhances query efficiency under high-load edge scenarios through dynamic task migration and resource-aware scheduling. The core contributions are threefold:

\begin{itemize}
    \item \textbf{Collaborative Scan Operator Design:} We introduce the first collaborative scan operator that dynamically offloads edge pressure based on runtime load sensing. This mechanism reduces query latency by 52.1\%–130.8\% in high-load conditions.
    
    \item \textbf{Cross-Layer Consistency Framework:} We build a multi-layer consistency mechanism combining hierarchical data localization and stream-based synchronization to eliminate redundant transmissions. An asynchronous state compensation strategy ensures zero-error migration under network fluctuations.

    \item \textbf{Adaptive Transmission Optimization:} We design a pipe-aware control mechanism tailored to query characteristics, which reduces network overhead by 43\%–79\%.
\end{itemize}

This work lays the foundation for CED collaborative query processing in time-series databases. Future research will focus on the following directions:
\begin{itemize}
    \item \textbf{Operator Extension:} Generalize the collaborative execution paradigm to additional operators such as join and last, addressing the demand for low-latency temporal correlation analysis across heterogeneous devices.
    \item \textbf{Intelligent Scheduling Enhancement:} Integrate a reinforcement learning-based dynamic scheduler into the \textit{ConfigNode} component to enable real-time load prediction and adaptive pipe control, thus improving resource utilization efficiency.
    \item \textbf{Hybrid Workload Optimization:} Develop a query-type-aware CED resource allocation model to optimize collaborative execution across compute-intensive and bandwidth-sensitive workloads.
    \item \textbf{Cache Strategy Optimization:} Enhance the cloud \textit{DataNode} caching algorithm to improve cache hit rates and accelerate repeated query performance.
\end{itemize}

\onecolumn
\twocolumn
\bibliographystyle{IEEEtran}
\bibliography{bibliography}

\begin{thebibliography}{10}
\providecommand{\url}[1]{#1}
\csname url@samestyle\endcsname
\providecommand{\newblock}{\relax}
\providecommand{\bibinfo}[2]{#2}
\providecommand{\BIBentrySTDinterwordspacing}{\spaceskip=0pt\relax}
\providecommand{\BIBentryALTinterwordstretchfactor}{4}
\providecommand{\BIBentryALTinterwordspacing}{\spaceskip=\fontdimen2\font plus
\BIBentryALTinterwordstretchfactor\fontdimen3\font minus \fontdimen4\font\relax}
\providecommand{\BIBforeignlanguage}[2]{{%
\expandafter\ifx\csname l@#1\endcsname\relax
\typeout{** WARNING: IEEEtran.bst: No hyphenation pattern has been}%
\typeout{** loaded for the language `#1'. Using the pattern for}%
\typeout{** the default language instead.}%
\else
\language=\csname l@#1\endcsname
\fi
#2}}
\providecommand{\BIBdecl}{\relax}
\BIBdecl

\bibitem{b2}
J.~Moon, S.~Cho, S.~Kum, and S.~Lee, ``Cloud-edge collaboration framework for iot data analytics,'' in \emph{2018 International Conference on Information and Communication Technology Convergence (ICTC)}.\hskip 1em plus 0.5em minus 0.4em\relax IEEE, 2018, pp. 1414--1416.

\bibitem{b3}
N.~Abbas, Y.~Zhang, A.~Taherkordi, and T.~Skeie, ``Mobile edge computing: A survey,'' \emph{IEEE Internet of Things Journal}, vol.~5, no.~1, pp. 450--465, 2017.

\bibitem{b4}
W.~Shi, X.~Zhang, Y.~Wang, and Q.~Zhang, ``Edge computing: state-of-the-art and future directions,'' \emph{Journal of Computer Research and Development}, vol.~56, no.~1, pp. 69--89, 2019.

\bibitem{b5}
S.~Madakam, R.~Ramaswamy, and S.~Tripathi, ``Internet of things (iot): A literature review,'' \emph{Journal of Computer and Communications}, vol.~3, no.~5, pp. 164--173, 2015.

\bibitem{b6}
S.~Rhea, E.~Wang, E.~Wong, E.~Atkins, and N.~Storer, ``Littletable: A time-series database and its uses,'' in \emph{Proceedings of the 2017 ACM International Conference on Management of Data}, 2017, pp. 125--138.

\bibitem{b7}
P.~Grzesik and D.~Mrozek, ``Comparative analysis of time series databases in the context of edge computing for low power sensor networks,'' in \emph{Computational Science--ICCS 2020: 20th International Conference, Amsterdam, The Netherlands, June 3--5, 2020, Proceedings, Part V 20}.\hskip 1em plus 0.5em minus 0.4em\relax Springer, 2020, pp. 371--383.

\bibitem{b8}
InfluxDB, \url{https://www.influxdata.com/}, n.d.

\bibitem{b9}
S.~N.~Z. Naqvi, S.~Yfantidou, and E.~Zim{\'a}nyi, ``Time series databases and influxdb,'' \emph{Studienarbeit, Universit{\'e} Libre de Bruxelles}, vol.~12, pp. 1--44, 2017.

\bibitem{b10}
C.~Wang, X.~Huang, J.~Qiao, T.~Jiang, L.~Rui, J.~Zhang, R.~Kang, J.~Feinauer, K.~A. McGrail, P.~Wang \emph{et~al.}, ``Apache iotdb: Time-series database for internet of things,'' \emph{Proceedings of the VLDB Endowment}, vol.~13, no.~12, pp. 2901--2904, 2020.

\bibitem{b11}
C.~Wang, J.~Qiao, X.~Huang, S.~Song, H.~Hou, T.~Jiang, L.~Rui, J.~Wang, and J.~Sun, ``Apache iotdb: A time series database for iot applications,'' \emph{Proceedings of the ACM on Management of Data}, vol.~1, no.~2, pp. 1--27, 2023.

\bibitem{b12}
{Apache IoTDB}, \url{https://iotdb.apache.org/}, n.d.

\bibitem{b13}
T.~Jiang, X.~Huang, S.~Song, C.~Wang, and J.~Wang, ``On tuning raft for iot workload in apache iotdb,'' in \emph{2024 IEEE 40th International Conference on Data Engineering (ICDE)}.\hskip 1em plus 0.5em minus 0.4em\relax IEEE, 2024, pp. 5307--5319.

\bibitem{b14}
{Apache IoTDB}, ``Data synchronization,'' \url{https://iotdb.apache.org/UserGuide/latest/User-Manual/Data-Sync_apache.html}, n.d.

\bibitem{b16}
G.~Graefe, ``Volcano-an extensible and parallel query evaluation system,'' \emph{IEEE Transactions on Knowledge and Data Engineering}, vol.~6, no.~1, pp. 120--135, 1994.

\bibitem{b17}
T.~Li, X.~Huang, J.~Wang, D.~Mao, Y.~Xu, and J.~Yuan, ``The design of apache iotdb distributed framework,'' \emph{Sci Sin Inform}, vol.~50, no.~5, pp. 621--636, 2020.

\bibitem{b18}
P.~Bonnet, J.~Gehrke, and P.~Seshadri, ``Towards sensor database systems,'' in \emph{International Conference on mobile Data management}.\hskip 1em plus 0.5em minus 0.4em\relax Springer, 2001, pp. 3--14.

\bibitem{b19}
O.~Diallo, J.~J. Rodrigues, and M.~Sene, ``Real-time data management on wireless sensor networks: A survey,'' \emph{Journal of Network and Computer Applications}, vol.~35, no.~3, pp. 1013--1021, 2012.

\bibitem{b20}
A.~Kanzaki, T.~Hara, Y.~Ishi, T.~Yoshihisa, Y.~Teranishi, and S.~Shimojo, ``X-sensor: Wireless sensor network testbed integrating multiple networks,'' \emph{Wireless sensor network technologies for the information explosion era}, pp. 249--271, 2010.

\bibitem{b21}
A.~G. Elias, J.~J. Rodrigues, L.~M. Oliveira, and B.~B. Zarpel{\~a}o, ``A ubiquitous model for wireless sensor networks monitoring,'' in \emph{2012 Sixth International Conference on Innovative Mobile and Internet Services in Ubiquitous Computing}.\hskip 1em plus 0.5em minus 0.4em\relax IEEE, 2012, pp. 835--839.

\bibitem{b22}
S.~K. Sarkar, \emph{Wireless sensor and ad hoc networks under diversified network scenarios}.\hskip 1em plus 0.5em minus 0.4em\relax Norwood, MA, USA: Artech House, 2012.

\bibitem{b23}
\BIBentryALTinterwordspacing
S.~Cui, H.~Wang, X.~Liu, Z.~Tian, and X.~Ding, ``Ts-cabinet: Hierarchical storage for cloud-edge-end time-series database,'' 2023. [Online]. Available: \url{https://arxiv.org/abs/2302.12976}
\BIBentrySTDinterwordspacing

\bibitem{b24}
S.~Prasad and S.~Avinash, ``Smart meter data analytics using opentsdb and hadoop,'' in \emph{2013 IEEE Innovative Smart Grid Technologies-Asia (ISGT Asia)}.\hskip 1em plus 0.5em minus 0.4em\relax IEEE, 2013, pp. 1--6.

\bibitem{b25}
M.~N. Vora, ``Hadoop-hbase for large-scale data,'' in \emph{Proceedings of 2011 International Conference on Computer Science and Network Technology}, vol.~1.\hskip 1em plus 0.5em minus 0.4em\relax IEEE, 2011, pp. 601--605.

\bibitem{b26}
T.~Beermann, A.~Alekseev, D.~Baberis, S.~Cr{\'e}p{\'e}-Renaudin, J.~Elmsheuser, I.~Glushkov, M.~Svatos, A.~Vartapetian, P.~Vokac, and H.~Wolters, ``Implementation of atlas distributed computing monitoring dashboards using influxdb and grafana,'' in \emph{EPJ Web of Conferences}, vol. 245.\hskip 1em plus 0.5em minus 0.4em\relax EDP Sciences, 2020, p. 03031.

\bibitem{b27}
C.~Wang, ``Research on storage methods of iot micro-service platform based on tdengine,'' in \emph{2023 3rd International Conference on Electronic Information Engineering and Computer Science (EIECS)}.\hskip 1em plus 0.5em minus 0.4em\relax IEEE, 2023, pp. 1113--1117.

\bibitem{b28}
B.~McBride and D.~Reynolds, ``Survey of time series database technology,'' 2020.

\bibitem{b29}
P.~Chen, W.~He, W.~Ma, X.~Huang, and C.~Wang, ``Iotdq: An industrial iot data analysis library for apache iotdb,'' \emph{Big Data Mining and Analytics}, vol.~7, no.~1, pp. 29--41, 2023.

\bibitem{b38}
D.~Lee, J.~Choi, J.-H. Kim, S.~H. Noh, S.~L. Min, Y.~Cho, and C.~S. Kim, ``Lrfu: A spectrum of policies that subsumes the least recently used and least frequently used policies,'' \emph{IEEE transactions on Computers}, vol.~50, no.~12, pp. 1352--1361, 2001.

\bibitem{b32}
J.~Xiao, Y.~Huang, C.~Hu, S.~Song, X.~Huang, and J.~Wang, ``Time series data encoding for efficient storage: A comparative analysis in apache iotdb,'' \emph{Proceedings of the VLDB Endowment}, vol.~15, no.~10, pp. 2148--2160, 2022.

\bibitem{b39}
T.~A. Li, X.~D. Huang, J.~M. Wang \emph{et~al.}, ``\BIBforeignlanguage{Chinese}{The design of apache iotdb distributed framework (in chinese)},'' \emph{\BIBforeignlanguage{Chinese}{Sci Sin Inform}}, vol.~50, pp. 621--636, 2020.

\bibitem{b33}
M.~Slee, A.~Agarwal, and M.~Kwiatkowski, ``Thrift: Scalable cross-language services implementation,'' \emph{Facebook white paper}, vol.~5, no.~8, p. 127, 2007.

\bibitem{b34}
R.~Srinivasan, ``Rpc: Remote procedure call protocol specification version 2,'' Tech. Rep., 1995.

\bibitem{b35}
R.~Liu and J.~Yuan, ``Benchmarking time series databases with iotdb-benchmark for iot scenarios,'' \emph{arXiv preprint arXiv:1901.08304}, 2019.

\bibitem{b36}
G.~Kumaran and V.~R. Carvalho, ``Reducing long queries using query quality predictors,'' in \emph{Proceedings of the 32nd international ACM SIGIR conference on Research and development in information retrieval}, 2009, pp. 564--571.

\end{thebibliography}

\end{document}